%% file: ms.tex
\documentclass[apj]{emulateapj}

\usepackage{natbib,enumerate,psfig,wasysym}
\newcommand{\msini}{\ensuremath{m \sin{i}}}
\newcommand{\feh}{\ensuremath{[\mbox{Fe}/\mbox{H}]}}

\newcommand{\persec}{\ensuremath{\mbox{s}^{-1}}}

\newcommand{\mjup}{\ensuremath{\mbox{M}_{\mbox{Jup}}}}

\def\astrosun {\mbox{$\odot$}}

\newcommand{\Msol}{\ensuremath{\mbox{M}_{\astrosun}}}
\newcommand{\Rsol}{\ensuremath{\mbox{R}_{\astrosun}}}
\shorttitle{MARVELS-1:  an Apparent Quintuple}
\shortauthors{Wright et al.}

\def\CEHW{1}
\def\PSU{2}
\def\Florida{3}
\def\Yale{4}
\def\Harvard{5}
\def\Wash{6}
\def\ND{7}
\def\OSU{8}
\def\Vandy{9}
\def\Brazil{10}
\def\LIneA{11}
\def\Canarias{12}
\def\Laguna{13}
\def\Fisk{14}
\def\Oklahoma{15}
\def\Valongo{16}
\def\Fisica{17}
\def\UVa{18}

\begin{document}

\title{MARVELS-1: A face-on double-lined binary star masquerading as a
  resonant planetary system; and consideration of rare false positives in radial velocity planet searches}

\author{Jason T.\ Wright\altaffilmark{\CEHW,\PSU},
Arpita Roy\altaffilmark{\CEHW,\PSU}, 
Suvrath Mahadevan\altaffilmark{\CEHW,\PSU},
Sharon X.\ Wang\altaffilmark{\CEHW,\PSU},
Eric B.\ Ford\altaffilmark{\Florida},
Matt Payne\altaffilmark{\Florida,\Harvard},
Brian L.\ Lee\altaffilmark{\Florida,\Wash}, 
Ji Wang\altaffilmark{\Yale}, 
Justin R.\ Crepp\altaffilmark{\ND},
B.\ Scott Gaudi\altaffilmark{\OSU}, 
Jason Eastman\altaffilmark{\OSU},
Joshua Pepper\altaffilmark{\Vandy},  
Jian Ge\altaffilmark{\Florida}, 
Scott W.\ Fleming\altaffilmark{\CEHW,\PSU},
Luan Ghezzi\altaffilmark{\Brazil,\LIneA},
Jonay I.\ Gonz\'alez-Hern\'andez\altaffilmark{\Canarias,\Laguna},
Phillip Cargile\altaffilmark{\Vandy},
Keivan G.\ Stassun\altaffilmark{\Vandy,\Fisk},
John Wisniewski\altaffilmark{\Oklahoma},
Leticia Dutra-Ferreira\altaffilmark{\Valongo,\LIneA},
Gustavo F.\ Porto de Mello\altaffilmark{\Valongo,\LIneA},
M\'arcio A.\ G.\ Maia\altaffilmark{\Brazil,\LIneA},
Luiz Nicolaci da Costa\altaffilmark{\Brazil,\LIneA},
Ricardo L.\ C.\ Ogando\altaffilmark{\Brazil,\LIneA},
Basilio X.\ Santiago\altaffilmark{\Fisica,\LIneA},
Donald P.\ Schneider\altaffilmark{\CEHW,\PSU},
Fred R.\ Hearty\altaffilmark{\UVa}
}

\email{jtwright@astro.psu.edu}
\altaffiltext{\CEHW}{Center for Exoplanets and Habitable Worlds, 525 Davey Laboratory, The Pennsylvania State University, University Park, PA 16802, USA}
\altaffiltext{\PSU}{Department of Astronomy and Astrophysics, The Pennsylvania State University, 525 Davey Laboratory, University Park, PA 16802, USA}
\altaffiltext{\Florida}{Department of Astronomy, University of
  Florida, 211 Bryant Space Science Center, Gainesville, FL,
  32611-2055, USA}
\altaffiltext{\Yale}{Department of Astronomy, Yale University, New Haven, CT 06511 USA}
\altaffiltext{\Harvard}{Harvard-Smithsonian Center for Astrophysics,
  60 Garden St., Cambridge, MA USA}
\altaffiltext{\Wash}{Department of Astronomy, University of Washington, Box 351580, Seattle, WA 98195-1580, USA}
\altaffiltext{\ND}{Department of Physics, 225 Nieuwland Science Hall,
 University of Notre Dame, Notre Dame, IN 46556-5670, USA}
\altaffiltext{\OSU}{Department of Astronomy, The Ohio State University, 140 West 18th Avenue, Columbus, OH 43210, USA}
\altaffiltext{\Vandy}{Department of Physics and Astronomy, Vanderbilt
  University, Nashville, TN 37235, USA}
\altaffiltext{\Brazil}{Observat\'{o}rio Nacional, Rua General Jos\'{e}
  Cristino, 77, Rio de Janeiro, RJ, Brazil, 20921-400}
\altaffiltext{\LIneA}{Laborat\'{o}rio Interinstitucional de
  e-Astronomia (LIneA), Rua General Jos\'{e} Cristino 77, Rio de Janeiro,
  RJ 20921-400, Brazil}
\altaffiltext{\Canarias} {Instituto de Astrof{\'\i}sica de Canarias
  (IAC), E-38200 La Laguna, Tenerife, Spain}
\altaffiltext{\Laguna}{Dept.\ Astrof{\'\i}sica, Universidad de La
  Laguna (ULL), E-38206 La Laguna, Tenerife, Spain}
\altaffiltext{\Fisk}{Department of Physics, Fisk University, 1000 17th Ave.\ N., Nashville, TN 37208, USA}
\altaffiltext{\Oklahoma}{HL Dodge Department of Physics \& Astronomy, University
of Oklahoma, 440 W Brooks St, Norman, OK 73019 USA}
\altaffiltext{\Valongo}{Observat\'{o}rio do Valongo, Universidade Federal do Rio de Janeiro, Ladeira do Pedro Ant\^{o}nio, 43, CEP: 20080-090, Rio de Janeiro, RJ, Brazil }
\altaffiltext{\Fisica}{Instituto de F\'{i}sica, UFRGS, Caixa Postal 15051, Porto Alegre, RS 91501-970, Brazil }
\altaffiltext{\UVa}{Department of Astronomy, University of Virginia, 530 McCormick Road, Charlottesville VA, 22901, USA}

\begin{abstract}
We have analyzed new and previously published radial
velocity observations of \mbox{MARVELS-1}, known to have an ostensibly
substellar companion in a $\sim 6$-day orbit.  We find
significant ($\sim 100$ m \persec) residuals to the best-fit model for the
companion, and these residuals are na\"ively consistent with an
interior giant planet with a $P = 1.965$d in a nearly perfect 3:1
period commensuribility ($|P_b/P_c -3| < 10^{-4}$).    We have
performed several tests for the 
reality of such a companion, including a dynamical analysis, a
search for photometric variability, and a hunt for contaminating
stellar spectra.  We find many reasons to be critical
of a planetary interpretation, including the fact that most 
of the three-body dynamical solutions are unstable.
We find no evidence for transits, and no
evidence of stellar photometric variability.  We have discovered two
apparent companions to \mbox{MARVELS-1} with adaptive optics
imaging at Keck; both are M dwarfs, one is likely bound, and the other is
likely a foreground object.  We explore false-alarm scenarios inspired by various curiosities in the data.  
Ultimately, a line profile and bisector analysis lead us to conclude that the
$\sim 100$ m \persec\ residuals are an artifact of spectral
contamination from a stellar companion contributing $\sim$ 15--30\% of
the optical light in the system.  We conclude that origin of this contamination is the
previously detected radial velocity companion to \mbox{MARVELS-1},
which is not, as previously reported, a brown dwarf, but in fact a G
dwarf in a face-on orbit. 
\end{abstract}
\keywords{stars: binaries: general --- stars: individual (TYC
  1240-945-1) ---
stars: low-mass, brown dwarfs}








\section{Introduction}

The measurement of precise radial velocities of stars has become a
crucial and standard part of the exoplanetary astronomer's toolkit,
and its application has now extended far beyond its early and most
successful application to well-studied, single, bright, chromospherically quiet
stars.  A push to apply radial velocity work to larger and thus fainter
samples will result in the discovery rare, touchstone systems that will inform
planet formation. 

The application to fainter stars in more crowded fields, as
required, for instance, to follow up candidate transiting planets
discovered photometrically, has necessitated attention to the
effects of rare and unlikely blend scenarios.  Since precise radial velocities
can both suffer from and reveal the nature of these blended 
systems, they are complementary to photometric
measurements and high spatial resolution imaging using adaptive optics, and thus serve a critical role in detecting, validating,
or ruling out planet candidates in a variety of situations
\citep[e.g.][]{Konacki03,Torres04,ODonovan07,Collier07,Santerne12,Kepler-62}.  Here, we present a case study of such an interplay in the case of an
insidious signal seen in the MARVELS survey for exoplanets.

\citet{Lee11} announced the detection of a short-period ``brown dwarf
desert candidate'' with minimum mass $28 \pm 1.5$
\mjup\ and period $P=5.89$ d orbiting the F star \mbox{TYC 1240-945-1} (\mbox{MARVELS-1}) as the first substellar companion discovered with
the Sloan Digital Sky Survey-{\sc iii} MARVELS Planet Search
\citep{Eisenstein11,Gunn06,Ge08}.
That work also reported analysis of precise radial velocities made with
the Hobby-Eberly Telescope (HET)
High Resolution Spectrograph using an iodine cell that had residuals to a one-companion fit
that were with surprisingly high ($\sim 100$ m \persec).  The magnitude of these residuals was attributed to systematic errors
resulting from the preliminary nature of the Doppler pipeline, which had not
been optimized for work at HET.

Herein, we report followup observations with HET using a
different pipeline that confirm the high residuals to a
single-companion fit and reveal that they are apparently consistent
with an inner companion, which we designate \mbox{MARVELS-1} $c^*$ with minimum 
mass of $\sim$0.8 \mjup\ (where the asterisk indicates the provisional
nature of the $c$ component).   \mbox{MARVELS-1} $c^*$, if real, would be unique among the known
exoplanets in the size and proximity of its larger companion.   While
the period of this putative inner companion is ambiguous because of aliasing
issues, the most likely solution is consistent with a perfect 3:1 period
commensuribility ($P= 1.965$ d), indicative of a mean-motion
resonance.   Such a system would be superlative in many ways, and
present a unique puzzle and opportunity for planet formation and
system evolution theorists.

We conduct an extensive examination of possible false
positives of increasing unlikelihood, and explore the reasons why the
MARVELS survey might be especially susceptible to them.  Following
Sagan's maxim that extraordinary claims require extraordinary
evidence, we adopt an attitude of healthy skepticism with respect to
the ostensible two-companion solution.  Ultimately, we find
that the known $\sim$6-day companion to \mbox{MARVELS-1} is in,
reality, another star in a nearly face-on orbit.

\subsection{Plan}

In Section~\ref{star} we describe our spectroscopic and imaging data
of the star \mbox{MARVELS-1}, including its imaged companions and the
basic stellar properties of all stars in the \mbox{MARVELS-1} system.

In Section~\ref{RV} we describe our precise radial velocity
measurements of \mbox{MARVELS-1} with data from HET and Keck
Observatories, our search for periodicities, and our best
double-Keplerian orbital solution.  

In Section~\ref{photometry} we describe our photometry of the system,
which shows no variability at the 0.2 mmag at the periods of
interest.  

In Section~\ref{dynamics} we describe our MCMC and dynamical analysis
of the system, which reveals only a few stable 2-companion solutions.

In Section~\ref{falsealarms} we discuss alternative explanations for
the signals we see.  Sections~\ref{FAP}--\ref{otherstars} describe a
variety of false alarm scenarios that we can rule out or seem
too unlikely for further consideration.  Section~\ref{pert} describes
a spectral contamination model that describes the data even better than the
double-Keplerian model in Section~\ref{RV}.  Sections~\ref{FAsummary}
and \ref{rarity} summarize our false alarm analysis and discusses the
discovery of rare systems in large planet searches such as MARVELS.

In Section~\ref{nomenclature} we discuss how we chose the names for
the objects in the \mbox{MARVELS-1} system, and we summarize our
findings in our concluding Section~\ref{conclusions}.  

\section{MARVELS-1}
\label{star}
\subsection{Basic Stellar Data}
\label{basics}
We have performed a reanalysis of the two FEROS spectra of \mbox{MARVELS-1}
presented by \citet{Lee11}, each of which have signal-to-noise ratios (SNR) of
340 and resolution $R\sim48,000$, and also analyzed 2 Apache Point
Observatory (APO) spectra
with $R\sim31,500$ and each with SNR of 210.  We have extracted basic stellar
parameters using three methods:  Spectroscopy Made Easy (SME)
\citep{SME}, {\sc StePar} \citep[][and references therein]{IAC}, and
BPG \citep[see Section 3.1.2 of][]{Wisniewski2012}.  This is similar to the method of
\citet{Wisniewski2012}, but now includes SME as implemented at Vanderbilt as a third pipeline.
We obtain excellent consistency in our estimates of $[{\rm
  Fe}/{\rm H}]$ and $T_{\rm eff}$

Averaging the six results (spectra from two sources, each through three
pipelines), we find $T_{\rm eff}=6297\pm
28$K, $\log{g} = 4.22 \pm 0.09$, $[{\rm Fe}/{\rm H}]= -0.13 \pm
0.04$, and $v_{\rm mic} = 1.50 \pm 0.03$ km \persec.  These results have a
significantly higher gravity 
\citep[a difference of 2-$\sigma$ from][]{Lee11}), indicating that \mbox{MARVELS-1} is in fact an F9 dwarf star,
not a subgiant as previously reported.  Applying the formulae of
\citet{Torres10} to these new values, we find $M_* = 1.25\pm0.06 \Msol$ and
$R_*=1.48^{+0.26}_{-0.22} \Rsol$ .

As we will see in later sections, there is likely $\sim$ 15--30\%
contamination in these spectra from a cooler companion, likely a G
dwarf, that may result in additional sources of systematic error in
these parameters.  

We have explored this possibility by analyzing the quality of the
spectral synthesis fits generated with Spectroscopy Made Easy
(SME) to extract parameters of the FEROS spectra.  The $\chi^2$
surfaces near minimum for these spectra have a parabolic character in
most dimensions, indicating a well-behaved fit and a lack of multiple
minima, as might be expected in contamination were a serious issue.  

We quantitatively explored the effects of spectral contamination on
our ability to extract precise stellar parameter by creating a
synthetic test using high-SNR APO spectra of HD 172051
($T_{\rm eff} = 5596\pm19{\rm K}, \log{g}=4.56\pm0.24$) and HD 22484 ($T_{\rm
  eff}=6063\pm19{\rm K}, \log{g}=4.29\pm0.16$). We shifted the spectra to have the same velocity shift and 
scaled them so that the former (cooler) star contaminated the latter with a
contribution of 30\%.  (We chose this level of contamination because it approximates
the flux ratio expected if these particular stars were at the same
distance, and is at the upper end of the contamination we expect in
MARVELS-1).

We then used the {\sc StePar} code to estimate the stellar parameters
of the blended spectrum and found $T_{\rm 
  eff}=5987\pm23$K, $\log{g}=4.44\pm0.19$.  These values are, not
surprisingly, intermediate to the correct values of the contributing
spectra. In the case of $T_{\rm eff}$ the value from the blended
spectrum is closer to the brighter star, and in the case of $\log{g}$
the value from the blended spectrum is consistent with that of the
brighter star within the quoted uncertainties.  

This test is not conclusive because the ``secondary'' star is more
metal poor than the primary by 0.2 dex ($\feh=-0.29\pm0.02$ vs.\
$-0.07\pm-0.02$), which may effect the final result (which yielded
$\feh=-0.11\pm0.05$), but it does suggest that the values we derive
here for \mbox{MARVELS-1} are not badly affected by a contaminating
spectrum.  Adequate caution is therefore needed when using
these parameters, which were calculated assuming an unblended spectrum.

\subsection{Companions Revealed by AO Imaging}

\subsubsection{Data acquisition and reduction}

We acquired images of \mbox{MARVELS-1} using NIRC2 (instrument PI: Keith Matthews) with
the Keck II adaptive optics (AO) system \citep{Wizinowich2000} on UT 24
June 2011.  Our initial set of observations consisted of dithered
images taken with the $K^\prime$ ($\lambda_c=2.12 \mu$m) filter.  Using
the narrow camera setting, to provide fine spatial sampling of the
NIRC2 point-spread function (PSF), we obtained nine frames with 4s of
on-source integration time each. A preliminary inspection of raw data
revealed the presence of an additional point source located 
0\farcs900  east of the
primary. Upon noticing the candidate companion, we obtained images
with the $J$-filter ($\lambda_c=1.25 \mu$m) to provide complementary
photometry, and detected the companion at 0\farcs897 separation,
consistent with the $K^\prime$ measurement.

We processed the images using standard techniques to replace
hot-pixels, flat-field the array, and subtract thermal background
radiation. After aligning and coadding the frames, we noticed a third
object in the field, to the north of the primary star and separated by
only  0\farcs152 in $K^\prime$ band (and $\sim$0\farcs18 in $J$ band,
though we adopt the $K^\prime$ band value because the PSF subtraction is
cleaner in those images).

\begin{figure}
\plotone{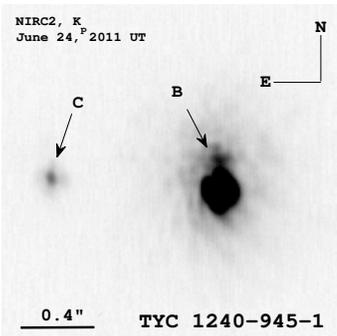}
\caption{$K^\prime$ band adaptive optics image from Keck/NIRC2 showing
  \mbox{MARVELS-1} B and \mbox{MARVELS-1} C.\label{Ji}}
\end{figure}

Figure~\ref{Ji} shows the image of \mbox{MARVELS-1} in the $K^\prime$
filter. Arrows indicate the location of the 0\farcs9 and 0\farcs15
companions. Inspecting the multi-color data set, we find that neither
object's position moves as a function of wavelength, demonstrating
that they are not speckles. Further, the PSF of each source is
qualitatively similar to that of the primary star.  

In the rest of this manuscript, we will refer to the primary star as
\mbox{MARVELS-1} A, the 0\farcs9 component as \mbox{MARVELS-1} C, and the 0\farcs15
component as \mbox{MARVELS-1} B (see Section~\ref{nomenclature} for an explanation). 

We performed aperture photometry to further characterize each point
source using the publicly available tool Starfinder
\citep{Diolaiti2000}. Often used for studying the Galactic center and
other 
crowded fields, Starfinder is optimized for AO photometry for which
numerous sources are spatially blended. By self-consistently fitting
each PSF, using an empirical iterative algorithm that minimizes
residuals, we were able to account for contaminating light from the
much brighter primary star to measure an accurate relative brightness,
angular separation, and position angle of each candidate in each
filter. Our astrometric and photometric measurements for both
companions relative to the primary are summarized in
Table~\ref{imaging}.   \mbox{MARVELS-1} C has colors comparable to an
M9V dwarf ($J-K=1.245 \pm$0.038), and \mbox{MARVELS-1} B has colors
($J-K=0.804\pm$0.071) consistent with spectral type K7V-M2V
\citep{Kraus07}.  

We have interpolated the absolute magnitude values in Table 5 of
\cite{Kraus07} for the spectral type of \mbox{MARVELS-1} A (F9V,
Section~\ref{basics}), and derive $M_K=$3.11 and $M_J=$3.42.  From the apparent 
magnitudes of \mbox{MARVELS-1} \citep[$K=9.032\pm0.017$ and $J=9.395\pm0.018$
according to][]{Lee11}, we derive a rough distance of $d=150\pm30$ pc from
spectroscopic parallax.  

\mbox{MARVELS-1} C is too bright to be an M9V dwarf at 
the same distance as \mbox{MARVELS-1} A, it is therefore likely an unbound
foreground star. The colors and absolute magnitude of \mbox{MARVELS-1} B are
consistent with the assumption that \mbox{MARVELS-1} A and B are at the same
distance.  

We obtain a consistent mass estimate for \mbox{MARVELS-1} B  of 0.62\Msol\
from the relations of \citet{Delfosse00} in both $J$ and $K$-band
(using $M_J=5.95$ and $M_K=5.20$). 

\mbox{MARVELS-1} has a proper motion of $\dot{alpha} = 33.6,
\dot{\delta} = -5.8$ mas yr$^{-1}$.  With
more than a one year time baseline from our first epoch observations,
future precision astrometric measurements using NIRC2, which has a
plate scale of $9.963 \pm 0.006$ mas \citep{Ghez2008} will be
sufficient to determine whether the two visual companions share a physical
association with the primary.  The angular proximity and the
consistency of the \mbox{MARVELS-1} B magnitudes and colors with being a
dwarf at the distance of \mbox{MARVELS-1} A argues that they are likely  bound.
  Irrespective of whether these sources represent a chance alignment along the line of sight, their flux
contribution must be accounted for when analyzing precise Doppler data.

\begin{deluxetable}{ccccc}
\tablecolumns{5} 
\tablewidth{0pt}
\tablecaption{Relative Astrometry and Photometry for the AO-resolved
  components of \mbox{MARVELS-1} with respect to the A component\label{imaging}} 

\tablehead{\colhead{component} &\colhead{$\Delta K^\prime$} &
  \colhead{$\Delta J$} & \colhead{Separation} & \colhead{PA}\\
& \colhead{mag.} & \colhead{mag.} & \colhead{arcsec.} & \colhead{deg.}}

\startdata 
B & 2.088$\pm$0.035 & 2.529$\pm$0.056 & 0.153$\pm$0.0017 & 2.6$\pm$0.33\\
C & 3.429$\pm$0.021 & 4.311$\pm$0.018 & 0.8998$\pm$0.0003 &86.25$\pm$0.02\\
\enddata
\end{deluxetable}

\label{AO}

\section{RV Observations}
\label{RV}
\subsection{HET Confirmation Data}
\citet{Lee11} reported nine precise radial velocity measurements made
with the Hobby-Eberly Telescope \citep{Ramsey98} High Resolution Spectrograph
\citep[HRS;][]{Tull98}  in late 2009 with Director's Discretionary Time as part of
the radial velocity confirmation program to the discovery of 
\mbox{MARVELS-1}{\it b} made with the MARVELS instrument.  These nine
observations were all acquired in 2009 December with an iodine absorption cell
calibrant \citep{Butler96b} at $R\sim60,000$.

These nine observations were reduced with a preliminary version of a precise
Doppler pipeline (provided by Debra Fischer).  This pipeline included
an instrumental profile model more appropriate for the slit-fed
Hamilton spectrograph at Lick Observatory and other components but was
not
optimized for work on HRS.

These nine points had significantly higher native precision than the
SMARTS and MARVELS velocities reported by \citet{Lee11}.  However these
velocity errors had to be scaled by a factor of $\sim 15$ to match the
observed $\sim 100$ m \persec\ scatter about the single-object Keplerian fit to the data
(though with a 6-component Keplerian fit to only 9 measurements,
the residual scatter was difficult to quantify precisely).
\citet{Lee11} attributed these high residuals to systematic errors in
their less-than-optimized Doppler pipeline.

\subsection{New Data}

The large residuals in the HET confirmation result did not appear completely random and
exhibited significant power at 0.5 d$^{-1}$.  This and a desire to
diagnose the remaining systematic errors in the Doppler pipeline led
us to request Director's Discretionary Time on HET to acquire
21 additional epochs on \mbox{MARVELS-1} at higher signal-to-noise
ratios.   These new observations
were made with the same spectrograph settings as the original nine observations. 
\label{aliasing}

Our new observations were made in late 2010 and early
2011, sufficiently long after the original observations to provide some
coverage of previously unexplored orbital phases. This system is
particularly challenging to observe at HET because the companion periods
are so close to integer numbers of sidereal days, and the fixed
elevation of HET requires that all observations occur 
within $\sim 1$ hour of two sidereal times (corresponding to the
rising and setting tracks of \mbox{MARVELS-1}).  This constraint made acquiring
good phase coverage problematic, and complicated our ability to rule out
competing and qualitatively different orbital solutions, and to
determine orbital eccentricities.   Nonetheless, we were able to adaptively
schedule our observations in the HET queue \citep{Shetrone07} so as to
optimally explore the 2 d period and eliminate all but one among many
competing orbital solutions near 2 d.

We also obtained a template and 14 velocity measurements from Keck using
the slit-fed HIRES spectrograph in August and September 2011.  These observations were reduced
using the usual methods of the California Planet Survey \citep[][and
references therein]{Howard10a,Johnson10,Wright11}.  Due to the
higher signal to noise ratio of these observations compared to the HET
data, the corresponding internal errors on the radial velocity
measurements are significantly smaller.

\subsection{Data Analysis: Template Analysis, Raw Reduction, and Doppler Pipelines}

We performed raw reduction of the echellograms from HRS using the
REDUCE package of \citet{REDUCE}.  This package required some
modifications to the slit function description to accommodate the
flat-topped nature of fiber-fed spectra.

\citet{Wang12} describes our Doppler pipeline.  In
brief, the code base developed by John
Asher Johnson and the California Planet Survey
\citep[e.g.][]{Howard09,Howard11a,Howard11b}
was modified for the HRS at HET.  This code is based on the principles of
\citet{Butler96b}.  We have demonstrated its precision on HRS 
spectra with the RV standard star ($\sigma$ Dra), for which we achieve
an r.m.s.\ scatter below 3 m \persec\ \citep{Wang11,Wang12}.

Our HET observations of \mbox{MARVELS-1} are not strictly analogous to our
$\sigma$ Dra observations, however, in that we have a lower
signal-to-noise ratio in both the template and radial velocity
observations.  To put a very conservative upper limit on the systematic errors inherent
in our Doppler pipeline on fainter targets, we analyzed three other
MARVELS targets, TYC 1194-144-1, TYC 3413-2471-1, and TYC 3410-1406-1,
in a manner identical to \mbox{MARVELS-1}.   

We stress that these
targets were observed at significantly lower signal-to-noise ratios
than \mbox{MARVELS-1} as exploratory science and so their RVs should have considerably
higher random and systematic noise.  These three stars represents our entire HET sample
of non-binary MARVELS targets, at present; we have no reason to
consider these to be RV stable stars, since these are, to our
knowledge, the first precise radial velocity measurements ever obtained on
these targets.   

\begin{figure*}
\plotone{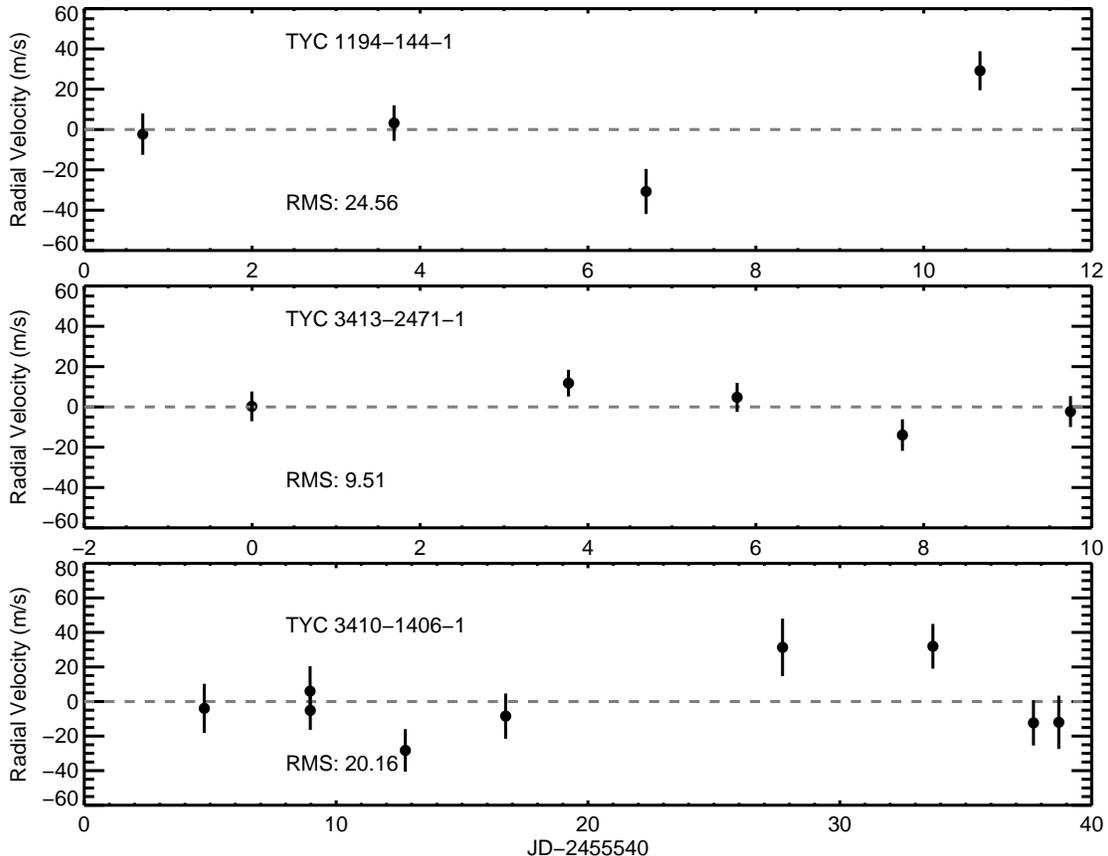}
\vspace{0.1in}
\caption{Radial velocities for three other MARVELS targets of similar
  apparent visual magnitude to \mbox{MARVELS-1}, but observed at lower signal-to-noise ratios
and analyzed using noisier template observations.  These observations
represent a conservative upper limit to any systematic errors in our
Doppler pipeline as applied in practice to typical MARVELS targets.
We have no a priori reason to believe that these stars are inherently
RV stable, and so the r.m.s.\ scatter of these radial velocities (listed in each pane in m \persec)
represent upper limits to our precision on these targets.  We expect
to achieve significantly better precision on \mbox{MARVELS-1}, closer to our
performance on $\sigma$ Dra, for which pipeline achieves 3 m \persec. \label{standards}} 
\end{figure*}

We present these three radial velocity targets in
Fig.~\ref{standards}.  The noisiest velocities here are for TYC
1194-144-1, with an r.m.s.\ dispersion of 25 m \persec; the best are for
TYC 3413-2471-1, at 9.5 m \persec.  Our \mbox{MARVELS-1} observations were
made at higher signal-to-noise ratios and we have no reason to suspect
\mbox{MARVELS-1} has levels of jitter this large; we thus expect to achieve
significantly better precision on \mbox{MARVELS-1}, closer to our performance
on $\sigma$ Dra.

We obtained an iodine-free stellar template at R=120,000 for this target, but the
signal-to-noise ratio was low due to the faintness of the star and the
slit losses at this high spectral resolution at HRS.  We opted
therefore to use the deconvolved stellar template obtained from Keck
observatory.  This template produced superior interal errors in the
HET velocities.

All of our HET and Keck velocities appear in Tables~\ref{vels} \& \ref{vels2}.  The reported
uncertainties are ``internal'' errors measured by the consistency with which different
portions of the spectrum (``chunks'') report the radial velocity of
the star \citep{Butler96b}.

\subsection{Keplerian fit}

We first performed a standard double-Keplerian fit, with no accounting
for dynamical (Newtonian) interactions or other sources of
non-Keplerian signal.

\begin{figure*}
\plottwo{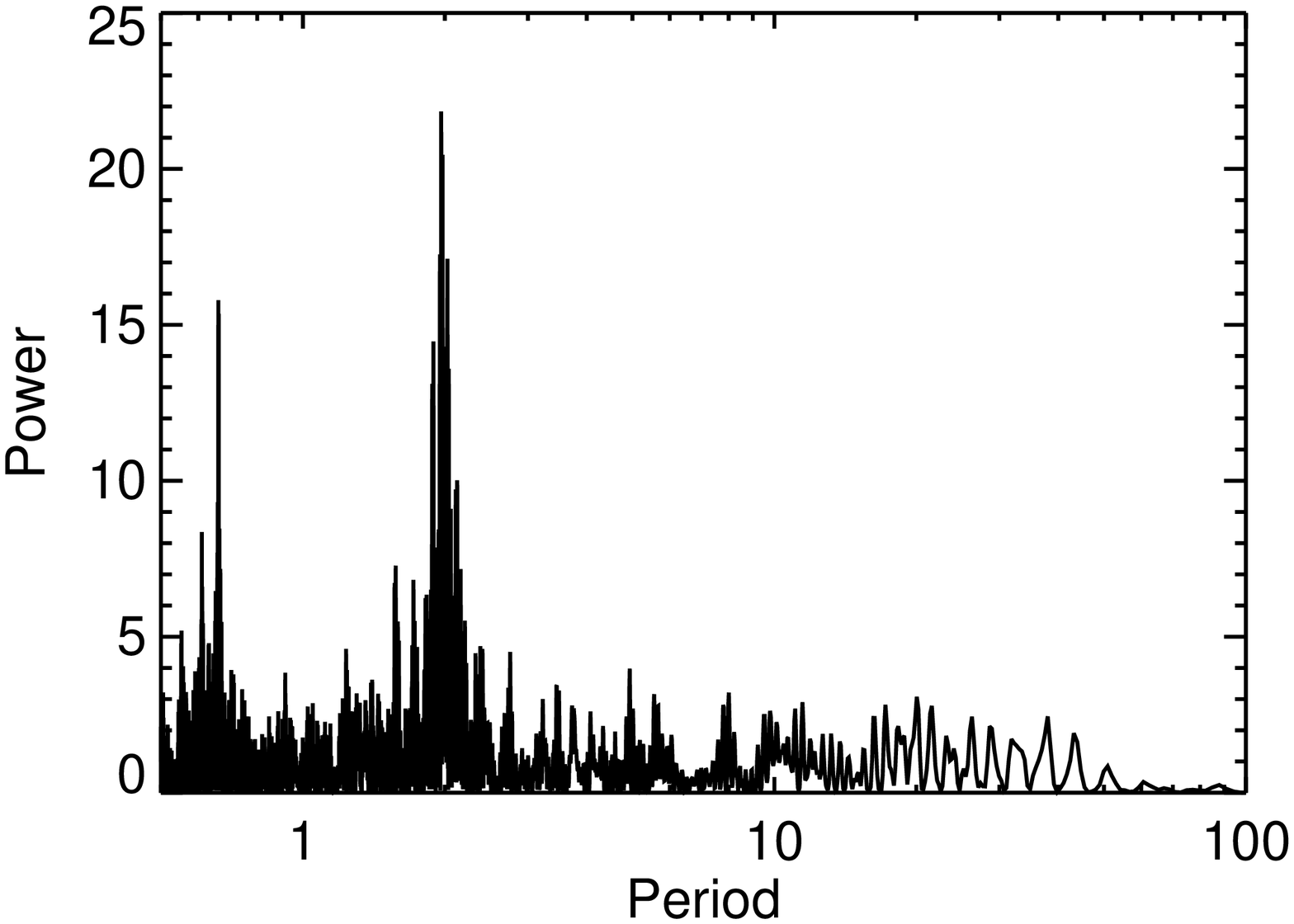}{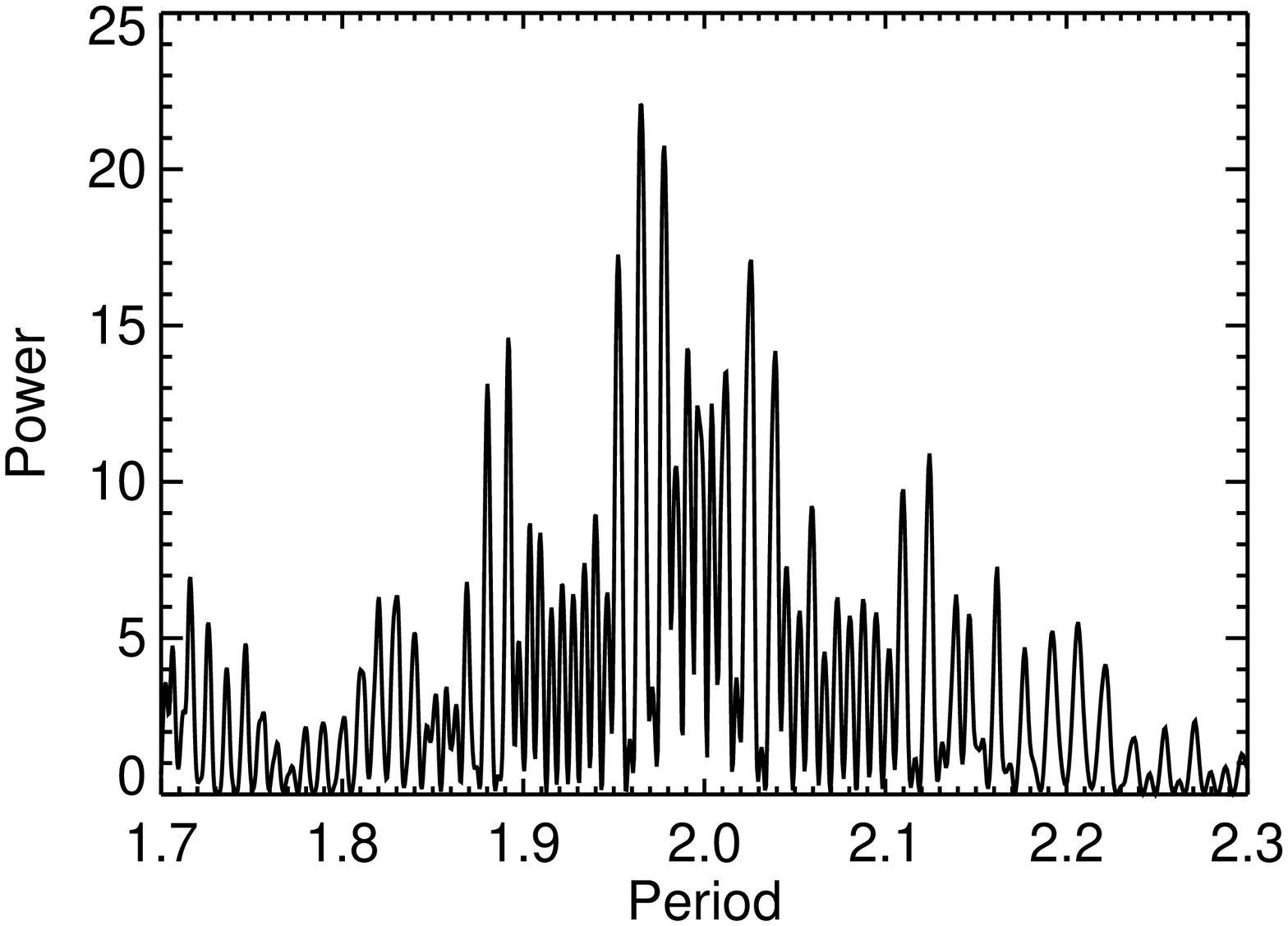}
\caption{{\it Left:} Periodogram of the residuals to a one-companion fit to the HET RV
  data.  There is substantial power near 2 days, but severe aliasing
  due to the near-integer period of the $c^*$ component produces a second
  peak near 0.66 days and large
  sidelobes to the tallest peaks.  The 2-companion
  Keplerian fit with the smallest residuals has $P_c = 1.965$ d. {\it
    Right:} Detail near 1.965 d.  \label{pergram}}
\end{figure*}

The $b$ component of the \mbox{MARVELS-1} system has such a large amplitude
that its orbital parameters are insensitive to the fit for the
$c^*$ component, and remain largely unchanged from the values in
\citet{Lee11}.    Fits for this system are complicated by the fixed
zenith angle of the HET (see \S~\ref{aliasing}) which makes it difficult to acquire good phase
coverage for this object, and produce significant side lobes in the periodograms of
residuals to a one companion fit, as shown in Fig.~\ref{pergram}.   

The periodogram shows two dominant peaks, near 2 d (0.5 d$^{-1}$) and
0.66 d (1.5 d$^{-1}$), which have similar amplitude and each with
substantial numbers of sidelobes.  These two peaks are aliases of each
other (their frequencies differ by exactly 1 sidereal day$^{-1}$).  As
we describe below, only the 1.965 day peak corresponds to orbital
solutions with residuals near our expectations; all of the other peaks
(including the 0.66 d peak) have best fits with at least 10 m \persec\ higher levels of residual
scatter.  Given that the 2 d peak has a perfect period
commensuribility with the primary signal and the 0.66 d does not, the
2 d peak is more likely to be the ``real'' signal.

\label{fit}

Regardless of its true period, the new observations appear to
strongly confirm the presence of a second periodic signal of apparent
semiamplitude $\sim 100$ m \persec\ for  \mbox{MARVELS-1}.  We have fit all 
velocity points using the RVLIN package of
\citet{Wright09b}, checking for a suite of orbital periods near 2 d
and 0.66 d to explore all of the peaks shown in Fig.~\ref{pergram}.  We
assumed 9 m \persec\ of jitter, chosen because it produced a  $\chi^2_\nu$ near 1 (the fit details are not strongly
sensitive to the amount of jitter assumed, and we performed a more
robust Markov-chain Monte Carlo (MCMC) calculation with jitter as a
parameter, as well, described 
in \S~\ref{PAYNE:METHOD}).  The parameters of this best 2-companion
fit are shown in Table~\ref{fittab}.  We calculated parameter
uncertainties for this fit using bootstrapping methods with the  {\sc
  boottran} routine \citep{Wang12}.
 
\begin{figure*}
\plottwo{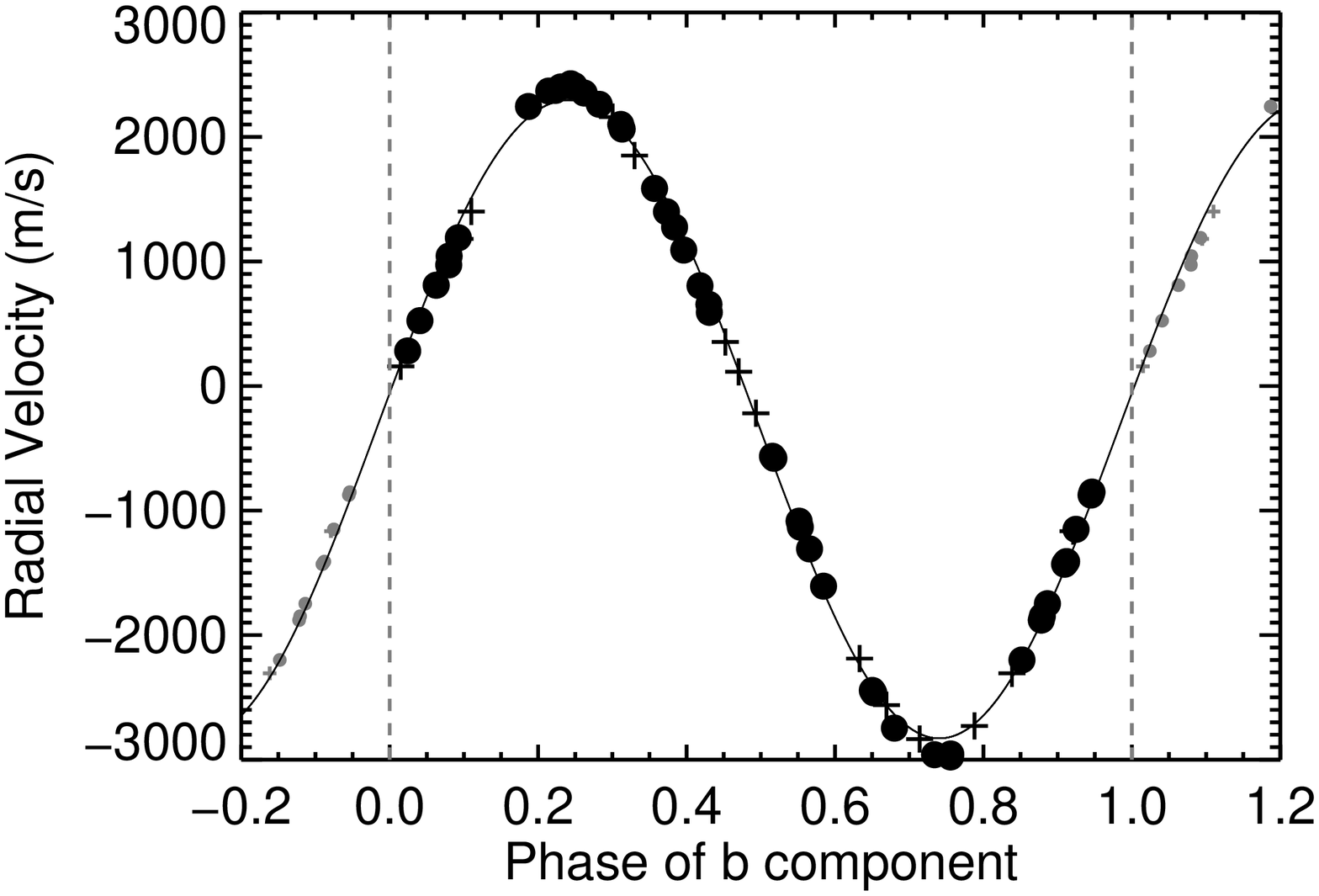}{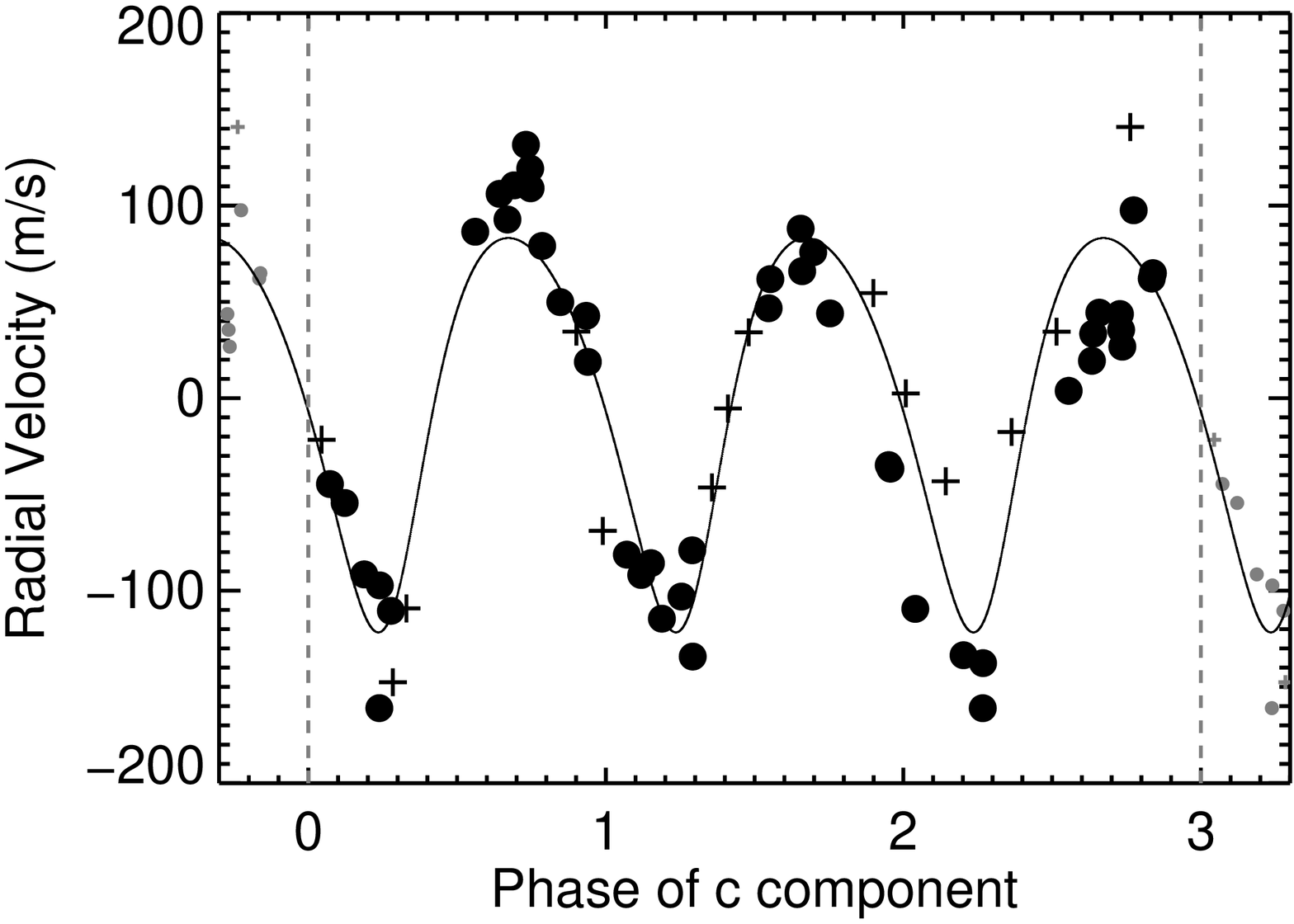}
\caption{Solid black disks are HET velocities, crosses are Keck
  velocities, and small gray symbols are repeated points.
  {\it Left:} Radial velocities phased at the period of the $b$
  component.  On this scale, the effects of the $c^*$ component are just
  visible. The internal (random) errors are smaller than the large
  plotted points.{\it Right:} Residuals phased to the period of
  \mbox{MARVELS-1} $b$ given in Table~\ref{fittab}, showing three consecutive 
  periods (only the gray points are repeated).  Deviations from a Keplerian
  curve are large, and appear to be a consistent function of phase of the
  $b$ component.  Note also that the zero crossings occur at the
  simultaneously with the zero crossings of the $b$ component.  We phase both
diagrams to have phase zero at the velocity zero with positive slope
of the primary component. \label{velplot}}
\end{figure*}

Figure~\ref{velplot} shows the measured velocities after subtracting the solution for the
$b$ component from the best 2-companion fit.  The temporal baseline of
the velocities span over a full year, giving coverage of most phases
of the orbit at this best fit period. The r.m.s. to the fit (the
standard deviation of the residuals, with no adjustment for the number
of model parameters) is 14.7 m \persec. 

Under the assumption that the $c^*$ component is real, we would
attribute this 9 m \persec\ of jitter needed to achieve $\chi^2_\nu$
near 1 to strong gravitational perturbations (see
\S~\ref{PAYNE:DYNAMICRESULTS}).

\begin{deluxetable}{rrr}
\tablewidth{0pt}
\tablecolumns{3} \tablecaption{HET Radial Velocities for
  \mbox{MARVELS-1}\label{vels}} 
\tablehead{\colhead{time} & \colhead{Velocity} & \colhead{Uncertainty} \\
 \colhead{BJD$-$2440000 (UTC)} &
  \colhead{m \persec} & \colhead{m \persec} }
\startdata 
\input{tab1.tex}
\enddata
\end{deluxetable}

\begin{deluxetable}{rrr}
\tablewidth{0pt}
\tablecolumns{3} \tablecaption{Keck Radial Velocities for
  \mbox{MARVELS-1}\label{vels2}} 
\tablehead{\colhead{time} & \colhead{Velocity} & \colhead{Uncertainty} \\
 \colhead{BJD$-$2440000 (UTC)} &
  \colhead{m \persec} & \colhead{m \persec} }
\startdata 
\input{tab2.tex}
\enddata
\end{deluxetable}

\begin{deluxetable}{cccc}
\tablewidth{0pt}
\tablecolumns{4}\tablecaption{Formal best-fit Keplerian Orbital Elements for Substellar Companions in the
 \mbox{MARVELS-1} System. Note that
  the parameters of our best model for the system appear in Table~7,
  not here.\label{fittab}}
\tablehead{\colhead{Property} &\colhead{$b$} & \colhead{$c^*$}}
\startdata
Per ($d$) &         5.895394(63) &       1.96492(29)\\
$T_0$ (JD-2440000) &  15498.059(91) &       15498.770(91)\\
e & 0.0160(16) & 0.134(56) \\
$\omega$ ($^\circ$) &  179.7(5.5) & 306(19) \\
$K$ (m \persec)& 2572.6(4.0)& 104.8(4.4) \\
$\msini$ (\mjup)& 27.6(1.5) & 0.764(42) \\
$a$ (AU) & 0.0702(19) & 0.03351(91) \\
\hline
r.m.s. (m \persec) & \multicolumn{2}{c}{14.7}\\
$\chi^2_\nu$ & \multicolumn{2}{c}{1.01}\\
jitter (m \persec) & \multicolumn{2}{c}{9}\\
$N_{\rm{obs}}$ & \multicolumn{2}{c}{30}
\enddata
 \tablecomments{This fit is dynamically unstable.  For succinctness, we express uncertainties using
   parenthetical notation, where the least significant digit of the
   uncertainty, in parentheses, and that of the quantity are to be understood
   to have the same place value.  Thus, ``$0.100(20)$'' indicates
   ``$0.100 \pm 0.020$'', ``$1.0(2.0)$'' indicates ``$1.0 \pm 2.0$'',
   and ``$1(20)$'' indicates ``$1 \pm 20$''.}  
\end{deluxetable}

\section{Photometry}

\label{photometry}

The transit search photometry of \citet{Lee11} using KELT data
\citep{KELT,KELT2} rules out photometric
variability on the period of either RV signal at the sub-mmag level; their Figure~3 illustrates that the star is photometrically constant.

\begin{figure*}
\plottwo{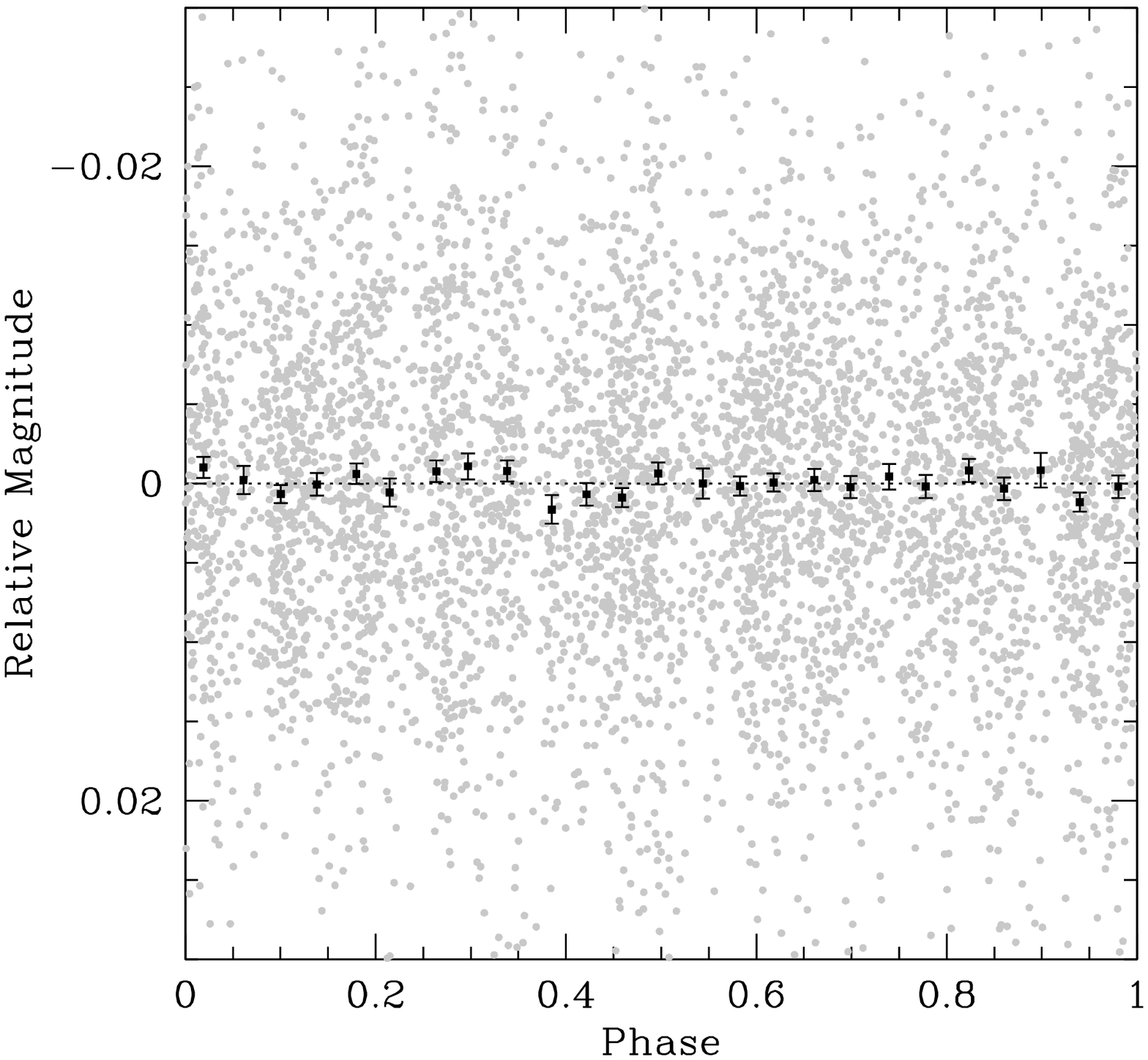}{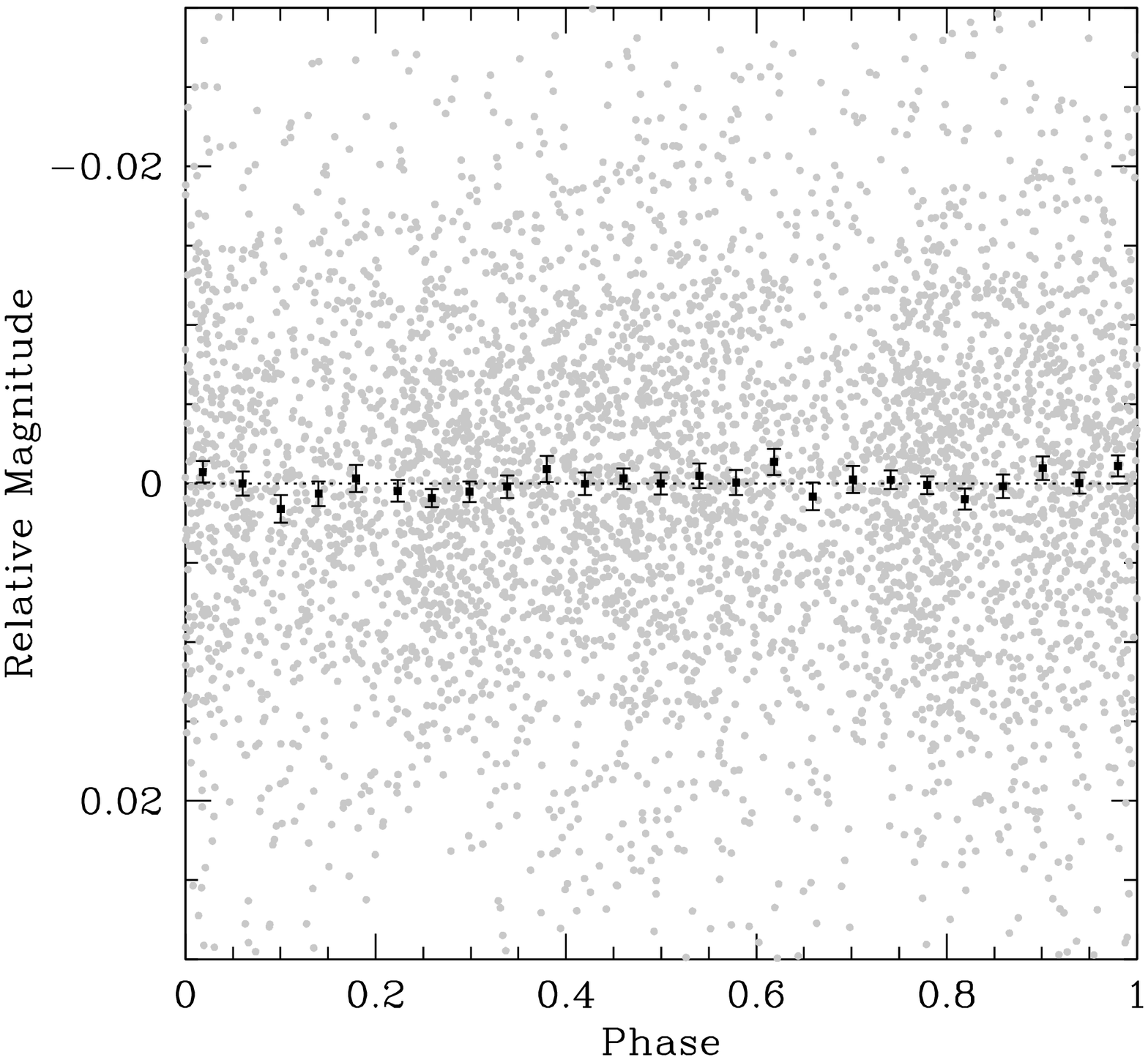}
\caption{KELT photometry of \mbox{MARVELS-1} phased at the periods of
  the $b$ (left) and $c^*$ (right) components.   Gray points show the
individual measurements and the black points show the data binned.
The binned data show that the star is photometrically stable at the
sub-mmag level, with no evidence of either transits or stellar pulsations.\label{phot}}
\end{figure*}

We have reanalyzed these data, and find that there is no evidence for
periodic variability, and the limits for P=1 to 10 days are quite
stringent.  The strongest peak in the periodogram in that interval has
$P=1.0218$d with amplitude $0.87\pm1.8$ mmag, and we associate this
peak with diurnal effects.  At the periods of the signals of \mbox{MARVELS-1} $b$
and $c^*$, the best-fit photometric amplitude is $ < 0.2$ mmag.  We
show the KELT data phased at these periods in Figure~\ref{phot}.

\citet{Lee11} used these data to search for transits of the $b$
component.  They were able to rule out transits with a depth of of
$\sim 0.2\%$ or greater, corresponding to companions of Jupiter radii
or larger for the original estimate of the radius of the primary.  If the 
$c^*$ component exists, its a priori transit probability is $\sim
20\%$, and for our new stellar radius, a Jupiter radius companion at
the period of the c$^*$ component would produce a transit with a
duration of $\sim 3$ hours and a depth of $\sim 0.5\%$.  For a strictly
periodic transit, we can robustly exclude such a transit signal.
However, we expect any transits of c$^*$ to exhibit large TTVs with
amplitudes of up to 10 hr due to dynamical interactions with $b$ (see
Section~\ref{PAYNE:DYNAMICRESULTS}).  Therefore, when phased to a
single period, the transit will get smeared out, thereby lowering the
signal-to-noise ratio with which we detect or exclude the signal.  

Nevertheless, we expect that such transits would probably still be
detectable in the binned curves, even
given the large possible transit timing variations: we can
approximate the signal as being diluted by a factor of the
duration/TTV amplitude, which is $\sim (3h/10 h)*0.5\% \sim
0.15\% $, a level similar to that for which \citet{Lee11} ruled out transits of
the $b$ component.

\section{MCMC and Dynamical Analysis}

\label{dynamics}
The novelty of this system demands a thorough analysis of the
orbital parameters of the components and their dynamics.  To this end,
we explored the space of two-companion orbital fits to the RV
data with a combination of MCMC analysis of the a posteriori
Keplerian orbital elements and with a dynamical analysis of some of
the families of solutions.  This MCMC analysis was performed at a
relatively early stage of the investigation, before the Keck data was
available. As such it is based on HET data alone. The MCMC analysis
was \emph{not} renewed following the Keck data observations, since
those data revealed that the $c^*$ signal was likely spurious.

The period ambiguity from the
aliasing of the period with the telescope observing constraints
prevent our MCMC code from settling on a unique period for the $c^*$
component, consistent with the heights of the sidelobes in
Fig.~\ref{pergram}.  We choose to focus on the 1.965 d
period, which has the lowest residuals and is favored by the data.  This inner
period is consistent with a perfect 3:1 PC (the best fit Keplerian
solution has $|P_b/P_c - 3| \sim 10^{-4}$).  
 
In summary, we find that most of the solutions in the MCMC chains are unstable on short
timescales, but there do exist stable solutions.  If the solutions we
explore here describe an actual qualitative behavior of the
system, then this system would exhibit significant perturbations, which
would imply that the osculating Keplerian orbital elements would vary
strongly and detectably on timescales of years, and that planetary
transits might be intermittent as the inclinations of the companions
oscillate.  Were transits to occur, they might exhibit large
($\sim 10$ hour) timing variations. 

\subsection{MCMC Methodology}\label{PAYNE:METHOD}
We analyzed the radial velocity measurements using a Bayesian framework following
\citet{Ford05b} and \citet{Ford06b}.  We assume priors that are
uniform in eccentricity, argument of pericenter, mean anomaly at
epoch, velocity zero-point and logarithm of the orbital period.
For the velocity amplitude ($K$) and jitter ($\sigma_j$), we adopted a prior
of the form $p(x)=(x+x_o)^{-1}[\log(1+x/x_o)]^{-1}$, with
$K_o=\sigma_{j,o}=1$ m \persec, i.e., high values are penalized
\citep[for a discussion of priors, see][]{Ford07b}. The likelihood for
radial velocity terms assumes that each radial velocity observation is independent and normally distributed about the true radial velocity with a
variance of  $\sigma_i^2+\sigma_j^2$, where $\sigma_i$ is the
published measurement uncertainty, and  $\sigma_j$ is a jitter parameter that accounts for additional scatter due to stellar variability, instrumental errors and/or inaccuracies in the model (i.e., neglecting perturbations or additional, low amplitude planet signals).

We used an MCMC method based upon Keplerian orbit fitting to calculate
a sample from the posterior distribution \citep{Ford06}. 
We use the algorithm described in \citet{Ford05b} to adaptively determine the appropriate step size to employ in our Markov chains, initializing the algorithm with system parameters drawn from the best fits found using the frequentist approach described in section 4. We calculated 8 Markov chains per MCMC realization, each with $\sim 2 \times 10^8$ states, and discarded the first half of the chains.
We calculated Gelman-Rubin
\citep{GelmanRubin} test statistics for each model parameter and
several ancillary variables and found no indications of
non-convergence amongst the individual chains (but see section
\ref{PAYNE:MCMCRESULTS} for discussion of the differences between
individual fits). 

Following Keplerian fitting procedure, we attempted n-body DEMCMC
fitting of the system \citep[using the method described
in][]{terBraak06,Payne11,Johnson11}, but found that either, (i) the total number of
observations is currently too small to constrain the parameter space
sufficiently for this numerically intensive mechanism to converge in a
reasonable time, or (ii) no self-consist planetary fit is possible for
this system.

To ensure that at least some fraction of the Keplerian fits are stable
in the strongly-interacting regime expected of objects that are so
close together and so massive, we took the results of the MCMC fits
and injected those systems into the Mercury $n$-body package
\citep{Chambers99} and integrated them forward for $\sim 10^5$ yrs
($\sim 10^7$ orbits). This exercise allowed us to distinguish the
long-term stable fits to the data from those systems which happen to
fit well but become unstable on very short timescales.  We define
``unstable'' in a loose manner: we reject any systems which undergo a
collision or change either of the planet's semi-major axes by $>50\%$. This
approach allowed significant variation in elements to occur, but rejects systems which qualitatively change their architecture. 

We assumed that all systems are coplanar and edge-on for the sake of
this analysis, hence all of the masses used in our n-body analyses are
\emph{minimum} masses. 

\subsection{MCMC Results}\label{PAYNE:MCMCRESULTS}
We performed a number of independent MCMC simulations, all of which
appeared to  converge individually.
However, the point of convergence often varied in different
runs, which manifested itself in different sets of solutions having
slightly different periods for the $c^*$ component (i.e.\ the chains
converged on local minima at the posterior probability distribution) essentially
corresponding to the sidelobes for the inner period evident in Fig.~\ref{pergram}.

For the rest of this analysis we concentrate on those chains that
converged on the local minimum characterized by the lowest values of
the jitter ($\sim 5$ m \persec).  These chains had an associated period for
the $c^*$ component of $P_c \sim  1.965$ d.  The chains which
converged to alternative local minima had significantly larger jitters
($\sim 10 - 50$ m \persec).  The prior for jitter is intentionally chosen to
have significant support even at realistically large jitters (which are highly
unlikely due to the absence of strong stellar activity indicators in
this star) so as to minimize the risk of overlooking interesting
structure in the goodness-of-fit surface.  

\begin{figure*}

\plotone{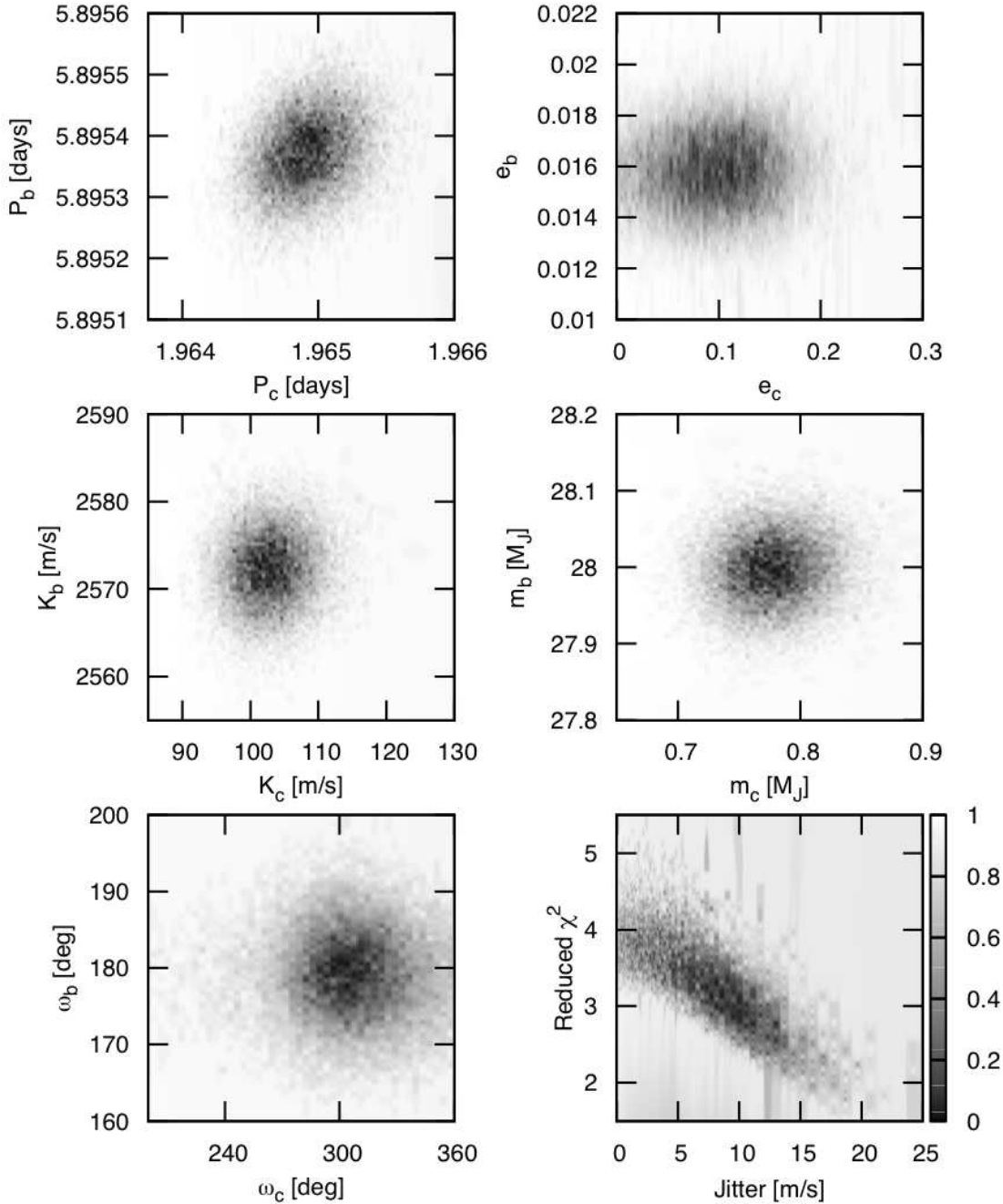}
\caption{The results from the Keplerian MCMC fitting algorithm
  which converged to low jitter solutions.  The top row contains
  $(P_b,P_c)$ and $(e_b,e_c)$, the second row $(K_b,K_c)$
  and $(m_b,m_c)$,  and the bottom row $(\omega_b,\omega_c)$ and (jitter,$\chi^2$).
The shading indicates the distance from the mode.  We find that
the period of the $b$ component is very tightly
constrained to be $P_{b}=5.895374_{-0.000067}^{+0.000066}$ d, while the $c^*$ component is less tightly constrained, lying in the range
$P_{c} = 1.96492^{+0.00029}_{-0.00028}$ d, where the uncertainties indicate $\sim$ 1-$\sigma$ confidence levels. 
}
\label{PAYNE:FIG:BESTFIT}
\end{figure*}

The results of our Keplerian MCMC fitting are presented in Fig
\ref{PAYNE:FIG:BESTFIT} and summarized in Table
\ref{PAYNE:TABLE:MCMC}. This analysis favors
$P_{c} = 1.96493^{+0.00030}_{-0.00028}$ d and
$P_{b}=5.895374^{+0.000067}_{-0.000066}$ d, implying a period ratio
that is consistent with exactly 
3:1, within the uncertainties ($P_b/P_c = 3.00029721^{+0.00046}_{-0.00049}$).  The
MCMC fits constrain the outer object to have a low eccentricity $e_{b} = 0.016^{+0.002}_{-0.002}$, but allows a large
spread in eccentricity for the $c^*$ component, $e_{c} =
0.10^{+0.06}_{-0.06}$.  

\begin{deluxetable}{ l  c  c c  }
\tablecolumns{4}
\tablewidth{0pt} 
\tablecaption{Orbital elements at JD 2,455,200 (noon 3 Jan 2010 UT) for the substellar
  companions resulting from the Keplerian MCMC analysis of \S
  \ref{PAYNE:MCMCRESULTS} in which jitter is included as a free
  parameter.  Note that
  the parameters of our best model for the system appear in Table~7,
  not here.\label{PAYNE:TABLE:MCMC}}

    \tablehead {Quantity & Mean & $+1\sigma$ & $-1\sigma$} 
    \startdata
$P_{b}$ (d)        & 5.89537      & 0.00007      & $-$0.00007     \\ 
$K_{b}$ (m \persec)        & 2572.4   & 4.3      & $-$4.4    \\ 
$e_{b}$        & 0.0158      & 0.0017      & $-$0.0017     \\ 
$\omega_{b}$ $(^\circ)$       & 179.4    & 6.0      & $-$6.1    \\ 
$M_{b}$ $(^\circ)$      & 202.3    & 6.0      & $-$6.0      \\ 
$m_b \sin{i_{b}}$ ($\mjup$)       & 28.01      & 0.05      & $-$0.05     \\ 
\hline
$P_{c}$ (d)        & 1.9649      & 0.0003      & $-$0.0003      \\ 
$K_{c}$ (m \persec)       & 103.1    & 5.0      & $-$4.7      \\ 
$e_{c}$        & 0.101      & 0.056      & $-$0.056         \\ 
$\omega_{c}$ $(^\circ)$        & 279.1    & 27.9     & $-$40.1     \\ 
$M_{c}$ $(^\circ)$        & 113.2    & 28.1     & $-$29.6     \\ 
$m_b \sin{i_{b}}$ ($\mjup$)       & 0.778      & 0.036      & $-$0.034      \\ 
\hline

Log Jitter     & 0.185      & 1.58      & $-$2.41    \\ 
$\chi^2$      & 3.711      & 0.527      & $-$0.52    \\ 
    \enddata
\tablecomments{Many of the solutions spanned by these
  parameters are dynamically unstable on short timescales.  }
  \end{deluxetable}


\subsection{Dynamical Analysis}\label{PAYNE:DYNAMICRESULTS}
%

As outlined in \S \ref{PAYNE:METHOD}, the results of the Keplerian
MCMC analysis in \S \ref{PAYNE:MCMCRESULTS} were used as input to the
Mercury n-body integrator \citep{Chambers99} to integrate the systems
for $10^5$ years.  ``Stepsize chaos'' can be a source of error in
numerical integrations, but it is negligible when the timestep $\Delta
t$ is smaller than the shortest physical timescale in the system by at
least a factor of 10 or 20 \citep{Rauch99}. Each integration was
carried out using a fixed timestep of $\Delta t = 0.01 P_c$, where
$P_c$ is the period of the inner planet. This exercise dramatically
reduces the number of suitable orbital fits, as $>99\%$ of the
solutions become unstable (using the approximate definition in  \S
\ref{PAYNE:METHOD}) over the course of the $10^5$ yr simulation.  

The stable systems favor lower periods and eccentricities for the
$c^*$ component, as shown in the first column of Table
\ref{PAYNE:TABLE:DYNAMIC} (we stress that the small number of 
systems ($\sim 20$) that remained stable means that the values quoted
suffer from small-number statistics). While the best fit values for
the period and eccentricity of the outer object are essentially
unchanged, the best fit values for the $c^*$ component become $P_{c} =
1.9644^{+0.0002}_{-0.0002}$ d and $e_c = 0.017^{+0.013}_{-0.014}$,
respectively. This means that the period ratio is increases 
slightly to $\sim 3.001$, but, more significantly, a much more circular
orbit for the $c^*$ component is strongly preferred (at least at the
epoch of first observation --- see below for further discussion).  

\begin{deluxetable*}{ l  c  c c  c  }
\centering
\tablecolumns{5}
\tablewidth{0pt}
\tablecaption{Orbital elements for the substellar companions resulting
  from integrations of the long-term stable results arising from the
  Keplerian MCMC analysis of \S \ref{PAYNE:MCMCRESULTS}.  Note that
  the parameters of our best model for the system appear in Table~7,
  not here.\label{PAYNE:TABLE:DYNAMIC}}
  
\tablehead {Quantity & Mean Value at Epoch\tablenotemark{1} &    1-Year Mean\tablenotemark{2}  & 1-Year Min\tablenotemark{2} & 1-Year Max\tablenotemark{2}}
\startdata
$P_{b}$ (d)        & $5.89534^{+0.00006}_{-0.00005}$      &       $5.944(1)$  &        $5.927(1)$ &        $5.961(1)$ \\
$a_{b}$ (AU)       & $0.0712776^{+0.0000005}_{-0.0000004}$        &       $0.071332(8)$    &        $0.071200(10)$   &        $0.071468(9)$ \\
$K_{b}$ (m \persec)      & $2572.4^{+3.4}_{-3.2}$       &      -    &       -   &       - \\
$e_{b}$        & $0.0168^{+0.0024}_{-0.0025}$ &       $0.020(2)$  &        $0.016(2)$ &        $0.024(2)$ \\
$\omega_{b} (^\circ)$        & $145.1^{+42.3}_{23.2}$       &       $176.5(6.3)$    &        $162.8(8.4)$   &        $189.1(5.9)$ \\[2pt]
\hline
$P_{c}$ (d)       & $1.9644^{+0.0002}_{-0.0002}$      &       $1.943(2)$  &        $1.904(3)$ &        $1.981(4)$ \\
$a_{c}$ (AU)         & $0.034097^{+0.000003}_{-0.000002}$   &       $0.033846(29)$    &        $0.033397(39)$   &        $0.034288(48)$ \\
$K_{c}$ (m \persec)       & $103.2^{+4.1}_{-4.4}$        &       -    &        -   &        - \\
$e_{c}$        &  $0.017^{+0.013}_{-0.014}$    &       $0.058(7)$  &        $0.0006(3)$       &        $0.14(1)$ \\
$\omega_{c}  (^\circ)$   &    $252.4^{+99.1}_{-192.8}$     &       $143.0(36.5)$   &        $0.0(1)$     &        $360.0(1)$ 
\enddata
\tablecomments{These elements span our favored dynamical solutions, however we cannot yet conclusively describe the true
  qualitative behavior of the system.  In addition, the perturbations in these
  solutions make these Keplerian osculating elements an inadequate
  description of the system.  See Table~\ref{fittab} for a note on
  error notation.}
   \tablenotetext{1}{Results for JD-2,455,200, averaged over the 20 results from the MCMC fitting of Table \ref{PAYNE:TABLE:MCMC} that were \emph{also} long-term stable.}
 \tablenotetext{2}{Extracted from detailed 1-year simulations
   initialized from the 20 individual starting conditions which
   underly the figures presented here in the ``Mean Value at Epoch''
   column, with data output at \emph{hourly} intervals and means, minimums, and maximums calculated.}
  \end{deluxetable*}

The results of the long-term stability analyses enumerated
in the first column of Table \ref{PAYNE:TABLE:DYNAMIC} should
\emph{not} be taken to be our final determination of the system
parameters. The low fraction of the systems from the Keplerian
fit which were stable indicates that while it is reassuring
that there exist stable solutions within the neighborhood of the
Keplerian fits, further work must be done to provide a reasonable
statistical sample of such orbits. The orbits are \emph{not} Keplerian
over the duration of the fits (see below for further discussion), so a
rigorous DEMCMC n-body fitting needs to be undertaken to generate such
a sample of stable orbits \citep[as outlined in][]{Johnson11}. As noted in \S \ref{PAYNE:METHOD}, our attempts at such an analysis were unsuccessful, as the routines did not converge. This could either be due to a lack of a sufficient number of high cadence observations, or may simply be due to it being impossible for a stable planetary system to give rise to the observed RV signal.

\begin{figure*}
%
\includegraphics[width=4.8in,angle=270]{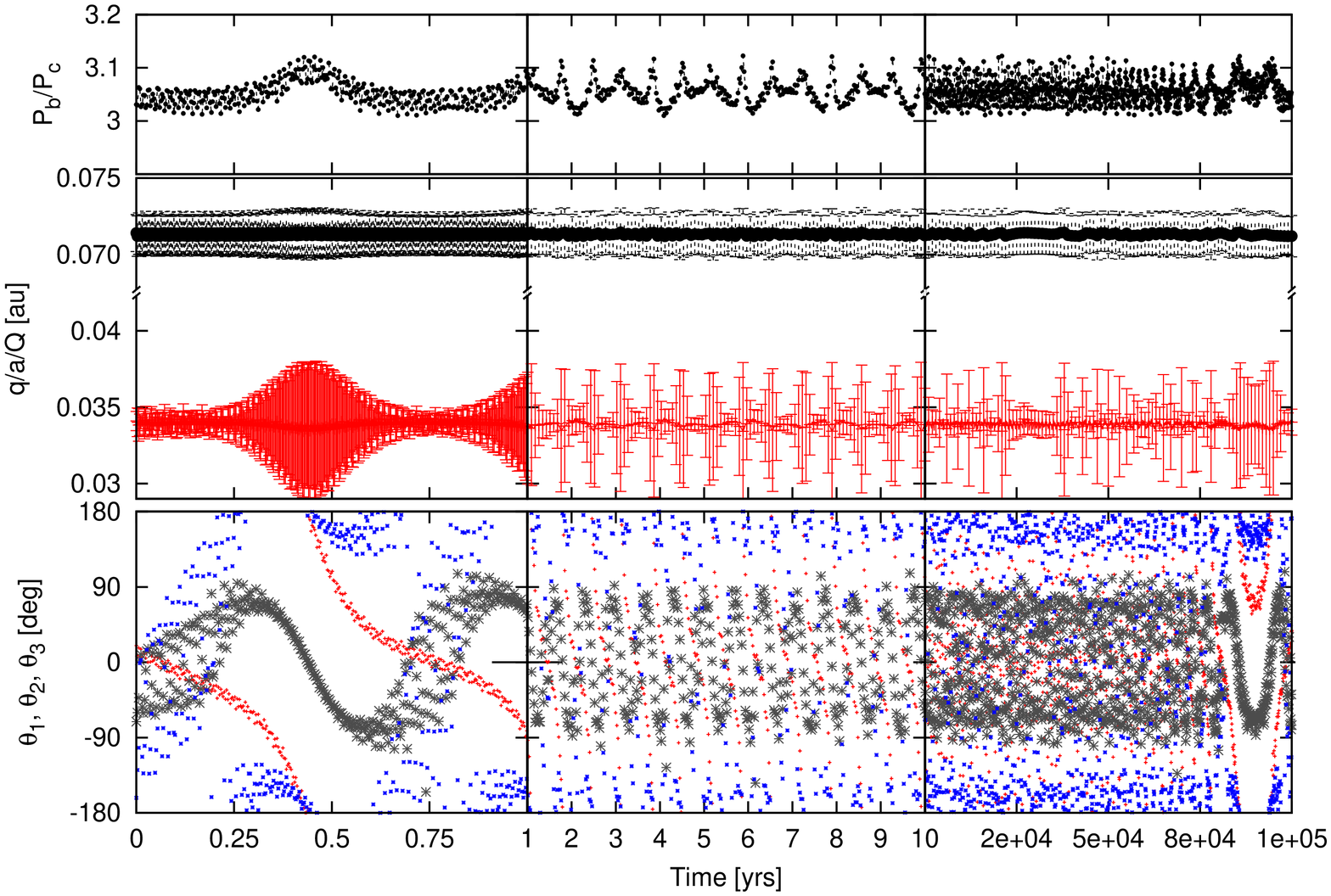}%
\caption{Orbital evolution for a sample long-term stable system.
We select one of the long-term stable systems from our dynamical analysis and plot in detail the evolution of the orbital elements over $10^5$ years, plotting on the left-hand side the detailed evolution over a single year, in the center the stretch of data from 1-10 years, on the right the long term evolution from 10 years on to $10^5$ years.
In the bottom panel we plot the three resonant arguments $\theta_1$ (red), $\theta_2$ (gray) \& $\theta_3$ (blue, see Eqn \ref{PAYNE:RESANG} for definitions), showing that at least one of the argument ($\theta_2$, gray) is librating, albeit with a relatively large amplitude.
In the second row we plot the range of orbital distances $q$
(pericenter, bottom of bars)
$a$ (semimajor axis, central line) and $Q$ (apocenter, top of bars) for the $b$ and $c^*$
components (black and red, respectively).  The eccentricity of the more
massive outer object changes very little, while the eccentricity of the $c^*$ component oscillates from $\sim 0.01 - 0.1$ over the course of $\sim 0.5 $ years. 
Finally, the characteristic semi-major axis oscillations in a resonant
system are present in this simulation (top row) and give rise to
variations in the period ratio $\sim 3\%$ on $\sim 0.5$ year
timescales (top plot).  If the $c^*$ component is real this result
raises the exciting possibility of detecting such period and
eccentricity oscillations directly from RV measurements.   
}
\label{PAYNE:FIG:EVOLUTION}
\end{figure*}


Fig \ref{PAYNE:FIG:EVOLUTION} shows the detailed evolution over the
course of $10^5$ yrs of various orbital elements in the system for one
of the stable solutions. In addition to the period ratio, pericenters, semi-major axes and
apocenters, we also plot the resonant argument $\theta_1, \theta_2$ and
$\theta_3$ (Murray \& Dermott 2000), defined as\footnote{where
  $\lambda$ is the usual mean longitude and we have taken the
  longitude of the ascending node $\Omega=0$ for simplicity.}
\begin{eqnarray}
\theta_{1} &=& 3\lambda_{b} - \lambda_{c} - 2\omega_{c}   \nonumber \\
\theta_{2} &=& 3\lambda_{b} - \lambda_{c} - (\omega_{c} + \omega_{b})    \label{PAYNE:RESANG} \\
\theta_{3} &=& 3\lambda_{b} - \lambda_{c} - 2\omega_{b}    \nonumber 
\end{eqnarray}
One can see that the system is in a resonant configuration, with
$\theta_2$ (gray asterisks) librating about $0^{\circ}$ with an
amplitude of $\sim 90^{\circ}$. One expects objects in resonance to
have semi-major axes which oscillate slightly as a function of time \citep{Murray99}, as demonstrated in the middle panel of Fig \ref{PAYNE:FIG:EVOLUTION}. Interestingly, in the case of \mbox{MARVELS-1} this leads to period oscillations of $\sim 3\%$ over the course of $\sim 200$ days ($< 40$ orbits of the outer body). Similarly, the eccentricity of the $c^*$ component in this sample simulation changes between $\sim 0.01$ and $\sim 0.1$ in $\sim 100$ d.

\emph{If} this example simulation were typical of the real system, then
such oscillations in period and eccentricity would be the largest and fastest seen to
date in any exoplanetary system.  This result would have several implications:

\begin{enumerate}[i]

\item For systems which interact strongly on short timescales,
  \citet{Laughlin01} and  \citet{Rivera01} demonstrated
  that Keplerian orbital fits are insufficient and that interaction
  terms need to be accounted for to fit the radial velocity
  data. While this is patently the case here for
  \mbox{MARVELS-1}, in this situation the interactions are so strong and orbital
  perturbations occur on such short timescales that one
  can only report the range, manner and timescale over which the
  quantity varies, in a manner similar to that undertaken for the
  proper orbital elements of asteroids \citep[e.g.][]{Knezevic02}.

\item The high masses and tight orbits of the bodies in TYC 1240 may
  provide an even more exquisite probe of a dynamically interacting
  system than the famed GJ 876 system \citep[][]{Marcy98,
    Delfosse98,Rivera10}: if a sufficient number of sufficiently precise RV
  observations can be taken with a sufficiently high cadence, then the dynamical interactions between
  the companions could reveal their inclinations and true masses. The
  advantage of the \mbox{MARVELS-1} system is that nearly a full
  cycle of the orbital element variations takes place within a single
  observing season, significantly boosting observational plausibility. 

\item Such strong interactions and associated large changes in orbital
  elements will likely drive significant transit timing variations
  (TTVs) in any hypothetical transit of the $c^*$ component.  

\item Our calculations were specifically for coplanar systems;
however, if the companions are mutually inclined, then the
precession of the orbit of the $c^*$ component could be sufficiently
strong and rapid that the system could quickly pass through episodes of
transit and non-transit, further complicating any transit search.

\end{enumerate}

We address point (i) above in Table \ref{PAYNE:TABLE:DYNAMIC}.  We
have taken all $\sim 20$ stable systems from our long-term
integrations and extracted the osculating elements at the epoch of the
first observation.  We report the mean and standard deviation of these
parameters across the 20 solutions in the first data column.  To
capture how they vary with time, we have also calculated the
mean and extremum values of these elements over the course of each of
the 20 simulations, and report the mean and standard deviation of these quantities
across the 20 simulations in the other three data columns.

For instance, over the period of the $n$-body calculations of the
stable solutions the eccentricity of the $c^*$ component has a mean value
of $e_c = 0.058$ and minimum and maximum values of $e_{c,min} =
0.0006$, $e_{c,max} = 0.14$, i.e., it changes from being almost
perfectly circular to having an eccentricity of 0.14. The standard
deviation figures are small, confirming that all 20 simulations
display such oscillations. 

We have conducted a preliminary investigation of the TTVs expected in
such a system, assuming that the system is edge-on to the
line-of-sight. We find that over the course of 1-year of observations,
TTVs of up to $\pm 10$ \emph{hours} can occur. Such large variation
would clearly be detectable, but would also have to be accurately
accounted for in any initial search to ascertain whether the system
transits at all (as any transit could easily be missed by many
hours).  

We conclude that there are some dynamically fascinating Newtonian
(interacting) interpretations of the observed radial velocity measurements,
but that most solutions are unstable.  This both adds to our
suspicions that the two-companion model for the observed signal is
incorrect and amplifies the extraordinary nature of the discovery if
it is real.

\section{Potential False Alarms}
\label{falsealarms}

At this point in our analysis, we have an intriguing and exciting result:  a
clear, strong signal suggesting a super-planetary system in a
near-perfect resonance with a strong likelihood of large and
detectable dynamical effects, and the potential for transits.  

But an application of a healthy skepticism reveals a few anomalies and
coincidences that require consideration.

First, 3:1 resonances are rare \citep{Wright11}.  Second,
such resonances are only perfect on average on long time scales;  the actual orbital
elements tend to librate about the equilibrium solution, so at any
given epoch the period commensuribility appears imperfect.  These
considerations make the apparent exactness of the commensuribility
suspicious.   Third, the fact that the data admit but a
few dynamically stable solutions (only 1\% of our simulations are
stable, and even those are highly dynamically active) suggests that
we may not have settled on the correct solution.

Further, the relative phases of the planets are suspiciously coincidental,
with both planets exhibiting radial velocity extrema and
zero-crossings simultaneously.  To understand why this must be a
coincidence, consider that for circular orbits these zero-crossings
correspond to inferior and superior conjunction (i.e.\ transit 
and secondary eclipse for edge-on systems).  As such, the
zero-crossings of the $b$ and $c^*$ components should only
appear simultaneous for a few special viewing angles.  

Finally, it seems inherently unlikely that an unusual system such
as this, for which there is no good analog among the thousands of
stars previously surveyed for planets, would be the first system to emerge from a
new planet survey (although, as we will see in Section~\ref{rarity}, the selection
criteria of MARVELS may make this not so unlikely as it seems).   

All of these oddities could be explained if the RV signal were not
actually the sum of two Keplerians, but some other shape that was
approximately fit by such a sum.  Could the faint, AO companions to \mbox{MARVELS-1}
have something to do with this unusual result?  

Prudence requires both ruling out other
explanations for the observed signal and understanding why MARVELS
might be especially sensitive to such systems.  In the face of the novelty of this
system and the suspicions raised by the coincidences above, healthy
skepticism demands that we rigorously consider all potential false
positives to confirm that each has truly been ruled out. 

\subsection{Formal False Alarm Probability}
\label{FAP}

The formal false alarm probability for the planetary companion is virtually
zero, due to its high amplitude compared to the r.m.s.\ residuals to
the fit.  We have confirmed this with a second-companion FAP calculation
using bootstrapping \citep[see ][ for a thorough
description.]{Wright08b}  We subtracted the best-fit 6 d period
Keplerian orbit from the velocities, and assumed the null hypothesis,
that the remaining scatter in the residuals was unpatterned noise of
some character (possibly non-Gaussian).  We then redrew these
residuals (with replacement), added to them the original 6-d signal, and performed a
thorough search of these artificial data for double-Keplerian
solutions.  We repeated this procedure to produce 1,000 trial data sets.
We recorded the r.m.s.\ value for the residuals for each trial, and compared
them with the residuals to the 2-Keplerian fit of the authentic data set.  

The sidelobes in the power spectrum of this data set makes searching for the
best 2-companion solution difficult, since our L-M fitter (RVLIN) is easily
trapped in local minima (i.e. sidelobes).  We therefore thoroughly
searched each bootstrapped dataset by testing the 10 most significant
peaks in the residual periodogram, sufficient to find the best fit in
the unscrambled data.  Even with this extra effort, 0 trials in 1000
had an r.m.s.\ residual as low as the actual set, consistent with the
amplitude of the true signal. 

This result demonstrates that the 2-day signal is real in the sense that it
is not a spurious periodogram peak introduced by the signal of the $b$
component and unstructured noise interacting with the window function
of the observations.  Demonstrating that the 
signal is due to a planet and not some spurious {\it astrophysical,
  instrumental}, or {\it methodological} false
positive requires eliminating alternative explanations (a task to which we
devote the rest of this section) and verifying
that the planet is physically plausible (i.e. that the derived orbit
is dynamically stable, which we did in \S~\ref{PAYNE:DYNAMICRESULTS}).

\subsection{Stellar Pulsations}
Practically speaking, any non-sinusoidal periodic radial velocity signature measured at finite signal-to-noise ratio can be well fit by a finite sum of sinusoids.  It is
therefore prudent to consider whether the entire 2.6 km s$^{-1}$
signal is in fact due to stellar oscillations, and the signal from the
apparent $c^*$ component is in fact merely the most important harmonic
component of a fundamental mode, or even a pulsational mode
of the star excited by resonant interaction with the brown
dwarf companion. 

\label{pulsing} 
The lack of photometric variation (see Section \ref{photometry}) would seem to rule out any pulsation
mechanism as the origin of the 2.6 km \persec\  signal; nonetheless we
have investigated the expected photometric variation of many
classes of variable stars, and conclude that the most important false
positive to consider for a star of this effective temperature and
periods near 1--10 d is that of a low-amplitude Cepheid (of either
type), such as Polaris.   

Polaris is well studied \citep[see, e.g.,][ for a summary]{Bruntt08}.
Its pulsational RV signature is sinusoidal, and its
amplitude has decreased from 6 km s$^{-1}$ to about 0.5 km s$^{-1}$ over
the last century.  Similarly, its photometric amplitude has decreased
from around 140 mmag to as low as 30 mmag in $V$ band in the year
2000, and as low as 10 mmag around 2005.  \citet{Bruntt08} find no
evidence of overtones of the 4 d pulsation period in RVs or
photometry.  \citet{Usenko05} find that the pulsational modes of
Polaris are not obvious from the spectra in that the effective
temperature of the star varies by no more than 100 K and does not
correlate with pulsational phase (although \citet{Polaris} find that the line bisector variations are
detectable when the RV amplitude is 1.5 km s$^{-1}$).

A low-amplitude Cepheid is thus a particularly insidious
false positive for a broad RV program, since it could in principle
have a small (and easily overlooked) $\sim 10$ mmag photometric
variation, small line bisector variation, essentially no
$T_{\rm{eff}}$ variation, and a large, easily detected $\sim 1$ km
s$^{-1}$ RV signal reminiscent of a hot Jupiter companion.  

There are several reasons this scenario, or a similar one, does not
apply to \mbox{MARVELS-1}.   
The first is that most secondary pulsation modes are not harmonics of the
dominant pulsation mode, but occur at a period with some
non-integer period ratio (beat Cepheids usually have
period ratios near 0.7 or 0.8, not 1/3).  The second is that our
prototype false positive, Polaris, shows no overtones at any period
ratio, integer or otherwise.  The third is that spectroscopic analysis of \citet{Lee11} conclusively shows that \mbox{MARVELS-1} is a
dwarf, not a giant or supergiant, and so is not expected to have pulsational modes with
periods near 6 d and in any case does not lie in the instability strip.  Finally, our
photometry in \S\ref{photometry} places upper limits on the
variability of \mbox{MARVELS-1} that are nearly
two orders of magnitude more stringent than the minimum variability of
Polaris, our prototype confounder.

\subsection{Stellar Activity and Star-Companion interactions}

While stellar activity can mimic the effects of planetary companions,
it is an inadequate explanation in this case.  \mbox{MARVELS-1} does not show
extraordinarily large Ca{\sc ii} H \& K emission or large $v\sin{i}$
values, and so we do not expect variations of order 150 m \persec, as
seen here for the $c^*$ component \citep{Wright05}, and certainly not at the level of the
signal of \mbox{MARVELS-1}$b$.  The stability of the photometry precludes
large spot-induced errors, and the low $v\sin{i}$ indicates that the
observed signals are likely at much shorter periods than the rotation period of
the star.  

The 3:1 period commensuribility is also inconsistent with
one of the signals being due to stochastic stellar activity, which
na\"ively should not occur at a harmonic of the orbital period of
\mbox{MARVELS-1}$b$.  

That said, the large mass and close separation of \mbox{MARVELS-1}$b$ suggests that we
consider interactions between it and its host star as a possible
source of unusual or novel effects on precise radial velocity
and bisector measurements.  Depending on the true mass and temperature of
\mbox{MARVELS-1}$b$, it may detectably heat the host star's surface,
altering its spectrum and convective flows there; it could cause a
prolate distortion of the host star; or it could perhaps resonantly
excite a normally dormant pulsation mode in the star at 1.965 d.   Any
of these effects could plausibly create a problematic RV signature
with correlated bisector changes.  

However, once again the photometric stability of \mbox{MARVELS-1}, documented in
\S\ref{photometry} at all of these frequencies, makes these
possibilities very unlikely.  Any variation in the output of
\mbox{MARVELS-1} of the sort described above strong enough to vary its
spectral lines' positions by 100 m\persec\ should also be apparent in
the photometry, at least at the level of our photometric precision.

Further, it is difficult to construct any such scenario that
produces significant power at three times the frequency of the $b$
component, but not significant power at twice the period (which, for
instance, prolate distortions would generate).

In short, any confounding effect along these lines would have to be previously
undetected by radial velocity surveys, generate large (100 m \persec) RV
variations at almost exactly three times the frequency of the $b$
component (and at no other frequencies), and do so with no detectable
photometric signature.  We can think of no such effects.

\subsection{Is MARVELS-1 $b$ a binary brown dwarf?}

One source of false positive is that \mbox{MARVELS-1} $b$ is actually an
equal mass binary brown dwarf with {\it total} minimum mass 28 \mjup,
in a 3:2 orbit-orbit resonance.  In this scenario, the variable
quadropole moment of the binary brown dwarf would induce a small
additional Doppler signature on \mbox{MARVELS-1} at twice the frequency of
the binary orbit, which we might observe as the 2-day signature.  The
binary would need to be nearly equal mass, or else the fundamental
orbital frequency would also produce a detectable signature.

Unfortunately, this particular false positive cannot be dispensed with
quickly, since the parameters are rather well constrained and the
predicted Doppler signature from such a scenario is similar to
that seen in the observations.  We present here a quick demonstration of
this fact, using a toy model.

If we assume circular, unperturbed orbits for both the binary and the binary's orbit
about \mbox{MARVELS-1}; denote the stellar mass $M$ and the mass of
each component of the binary $m (= 14 \mjup)$; denote the semimajor axis of the orbit of
 the center of mass of the binary around \mbox{MARVELS-1} $a$ (corresponding to period $P = 6$ d); and denote the total orbital separation $2d$ of the binary
components from each other; we can use Kepler's laws to derive a
rough, order-of-magnitude expression for the ratio $d/a$:

\[ \left(\frac{d}{a}\right)^3 = \frac{1}{9} \frac{m}{M+2m}\]






\noindent The semi-amplitude of the variation in the star's acceleration due to this
variable quadropole for small $d/a$ can be approximated as 






\begin{eqnarray}
\frac{K^\prime}{K} & = & \frac{3}{2}\left(\frac{d}{a}\right)^2\\
 & = & \frac{3}{2}\left[ \frac{1}{9}\frac{m}{M+2m} \right]^\frac{2}{3} 
\end{eqnarray}

\noindent where $K^\prime$ represents the semiamplitude of the
perturbative signal.  For $m/M \sim 4 \times 10^{-2}$ (assuming a high inclination for which the
total BD binary mass is actually 80 Jupiter), we have $K^\prime/K \sim 0.03$ or
one part in 25.  Our actual semiamplitude ratio from \S~\ref{fit} is one part in 22, which is
close enough that further analysis might be warranted.

Our healthy skepticism thus finds significant purchase here.  In order to claim
that the $c^*$ component is a separate planet and not a spurious
residual from the signal of a binary brown dwarf one must perform a more detailed
analysis and show that such a scenario is dynamically impossible.  We
did not perform such an analysis for this system here, because we have
a more important confounding effect to consider.

\subsection{RV Signals From Fainter Stars}
\label{otherstars}

An obvious potential source of problems is the presence of the two
companion stars described in \S\ref{AO}, whose spectra might
contaminate that of \mbox{MARVELS-1} and produce complex and spurious
velocity signatures.  

We might also suspect that the entire RV signature is due to
pulsations of one of the background objects, of the sort discussed in
\S\ref{pulsing}.  This explanation also suffers from the fact that the
photometry in \S\ref{photometry} detects no variability for this
system;  if the pulsation modes are so dilute as to be invisible in
our photometry they should not appear as a 2.6 km \persec\ signature
in the RVs .  This explanation also shares the difficulty that there are no
classes of pulsing stars that exhibit the 3:1 period commensuribility
we see.

One general source of false positives with blended targets is a faint
stellar binary whose signal is diluted by a brighter coincident star.
In this case the Doppler signature of the binary is proportionally
diluted, and the signal of the binary can be misinterpreted as a
planetary signal \citep{Torres04}.

We can rule out the 0\farcs9 companion, which lies near the edge
of our 2$^{\prime\prime}$ HRS fiber during nominal pointing.  Seeing,
tracking, and pointing variations should thus produce variable levels of contamination, and
thus a variable Doppler signature that would lack any particular
periodicity.  Also, the contamination should be at a completely
different level in our Keck velocities (since HIRES is slit-fed).  Since we see a consistent
signal in both the HET and Keck velocities, if the signal is due to
a binary, then it must be the binarity of the 0\farcs15 object or another
unresolved contaminant.  
 
The 3:1 period commensuribility further restricts the possibilities of
a blend.  If the source of the 100 m \persec\ amplitude signal is
to be attributed to the binary motion of one of the companion stars, we must accept the
coincidence that the effect has a frequency surprisingly near a small integer
multiple of the orbital frequency of \mbox{MARVELS-1}$b$.  Alternatively, if the
entire RV signature of \mbox{MARVELS-1} is to be due to a contaminating object, then
we still must confront the novelty of this short-period 3:1 resonant
system, except with significantly higher RV amplitudes (and thus
masses and luminosities) for the objects (since their signal would be significantly
diluted by \mbox{MARVELS-1}).   

In short, it is difficult to construct a plausible scenario whereby two signals
in a near-perfect 3:1 resonance are the product of anything other than
Doppler signals from the primary star, \mbox{MARVELS-1}. 

\subsection{Effects of spectral contamination}

\label{pert}
In this section, we consider other sources of spurious Doppler
signatures from spectral contamination.  In the discussion below we
will imagine how a cross correlation procedure would be corrupted by
the presence of a contaminating stellar spectrum.  We will refer to a ``template'' spectrum (in practice, taken
without the iodine cell in the beam) and an ``epoch'' spectrum (a
subsequent observation, through the iodine cell).  The change in
velocity measured for an ``epoch'' spectrum with respect to the
``template'' spectrum is measured as the velocity shift at which the
cross-correlation function (CCF) is maximized.  This peak occurs at
$\Delta v = v_e-v_t$, where $v_e$ is the velocity of the star at
epoch, and $v_t$ was the velocity of the star during the template observation.

If a companion star with radial velocity $v_c$ were contaminating our spectra, then this
contamination would produce a weak signal in the CCF at the velocity
shift that aligns it with the template spectrum, $\Delta v = v_c-v_t$ (this
effect is weak both because the spectral types do not match and
because the contaminating spectrum is weak.)  There would also be a
similar spurious peak from the contamination in the {\it template}
spectrum correlating with the primary spectrum in the epoch
observation, at $\Delta v = v_e-v_c$ (there is also a second-order
peak from the contaminating spectrum interacting with itself at
$\Delta v =0$.)

In this scenario, as the spectrum of \mbox{MARVELS-1} shifts due to
the influence of the brown dwarf, its spectrum will periodically
align with the contaminating spectrum.  When this happens, the
``true'' peak in the CCF at $\Delta v = v_e-v_t$ will align with the
spurious peak at $v_c-v_t$.   At values of $v_e$ near $v_c$, the two
CCF peaks will be blended, and the peak of their summed shape, which will
be asymmetric, will not be at $v_e-v_t$, but will rather be ``pulled''
towards $v_c-v_t$ by an amount that depends on the amount of
contamination and the magnitude of $v_c-v_t$.

Since the widths of these peaks are proportional to the line widths in
the spectra, this effect will only be important for values of
$|v_e-v_c|$ less than or similar to a typical line width, and the
effect should decrease in magnitude rapidly for larger values.

The second spurious peak (at $\Delta v = v_e-v_c$) does not produce a
similarly time-variable velocity anomaly; it will pull the
true CCF peak (at $\Delta v = v_e-v_t$) by a constant amount, which
will be subtracted as a zero-point offset in the resulting differential velocities.

The time-variable component of this ``peak pulling'' could thus cause
systematic errors in our measured velocities such that they are erroneously closer to the
velocity of the contaminating star when the spectra are near
alignment.  This would manifest as an anomalous plateau near the preferred velocity,
with a characteristic width determined by the typical line widths of
the primary and contaminating lines.  Since the amplitude of the RV
variation from \mbox{MARVELS-1} $b$ is of similar magnitude to the 
line widths of \mbox{MARVELS-1}, we would expect all of the velocities to be
affected to some degree (except at the preferred velocity where the
spectra are perfectly aligned), but most severe when the velocity
difference is greatest.  

Although the AO companions to \mbox{MARVELS-1} are too faint to
contribute sufficient flux in the optical for this effect to be
significant, we nonetheless explore it out of an abundance of caution.
Below, we build a model for this effect and apply it to our Keck and
HET velocities.

\subsubsection{A general model for ``peak pulling''}
\label{peakpulling} 

We have modeled such an anomaly as
$v_{\rm{pert}}$ where the velocities show a 
systematic tendency towards a preferred velocity $v_0$ when they are
within a characteristic line width $\Delta v$ of $v_0$:

\[v_{\rm{pert}} = k(e^{-((v-v_0)/\Delta v)^2}) (v_0-v) \label{vpert}\]

\noindent where $k$ is the magnitude of the effect.  

\begin{figure*}
\plottwo{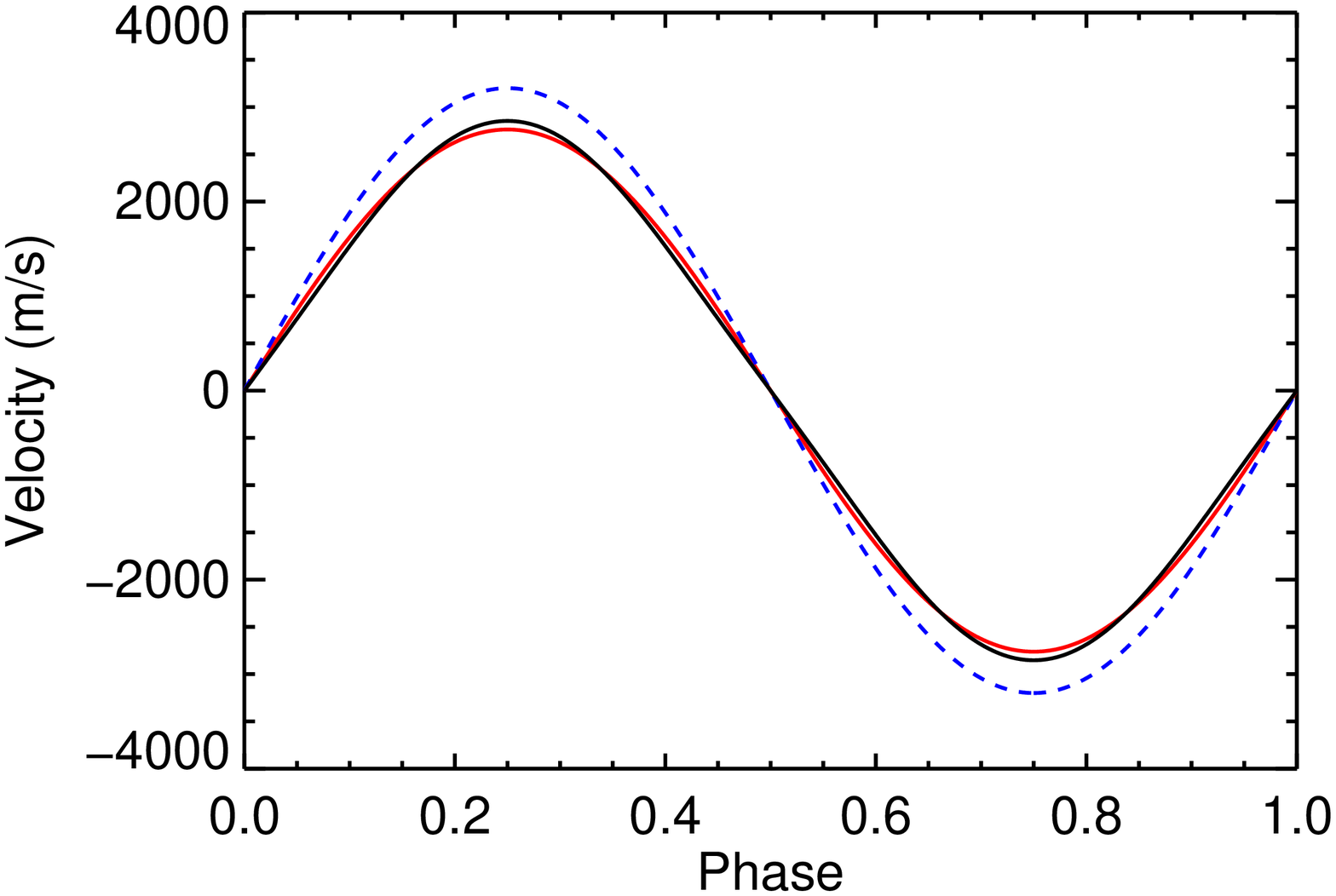}{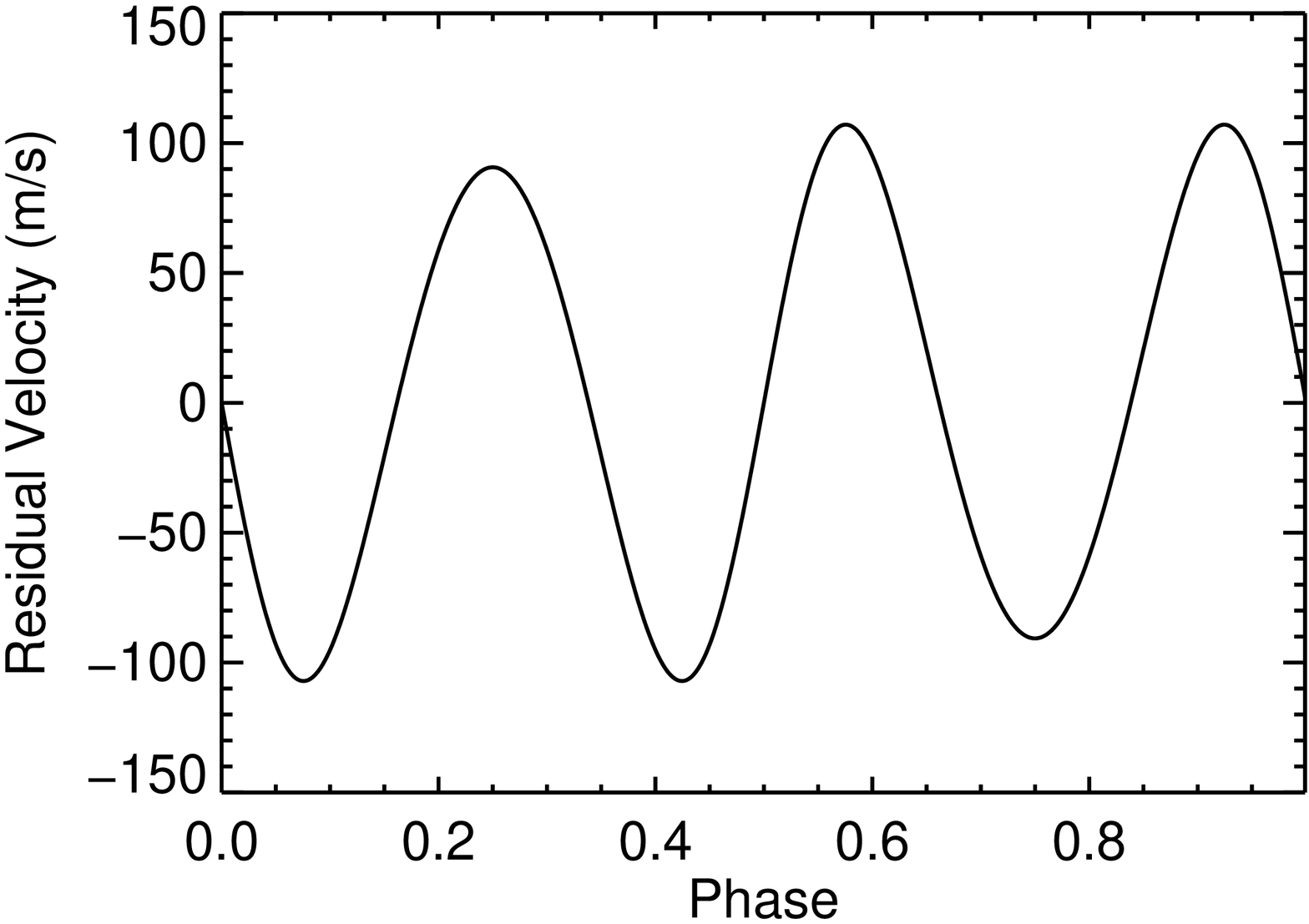}
\caption{Left: Velocities perturbed by a simple contamination model
  (black) and best circular fit (red).  The blue dashed line
  illustrates the true, unperturbed
  velocity.  Right: Residual velocities (perturbed minus fit) showing
  the nearly sinusoidal, triple-peaked residuals (i.e., black
  curve minus red curve from the left panel).  The
  shape of these residuals are sensitive to the details of the contamination model,
  but will show a perfect 3:1 period commensuribility for special
  values, like the ones we have chosen here.  We phase both
diagrams to have phase zero at the velocity zero with positive slope
of the primary component.
Compare these
  figures with Figure~\ref{velplot}.  \label{ouch}}
\end{figure*}

We have generated artificial data assuming $\Delta v =3.5$ km \persec,
$k=0.25$, and $v_0=0$ as a
perturbation on a sinusoidal signal of semiamplitude 3 km \persec, and
performed a sinusoidal best fit to the resulting velocities;  the
result is presented in Figure~\ref{ouch}.  Note the similarity of these
curves to those in Figure~\ref{velplot}.  

Here, our healthy skepticism and abundant caution seems to have paid off.  This ``peak
pulling'' scenario naturally produces a residual signal to a
sinusoidal fit whose dominant mode is at exactly 3 times the frequency of
the primary signal.  

This result can be understood from the symmetry produced
by a pulling near the velocity zero, which produces a signal that
shares the symmetries of a sinusoid, but in this case is slightly
taller and narrower.   As Figure~\ref{ouch} illustrates, the best-fit
sinusoid and perturbed velocity 
curves meet at the $v=0$ points (because $v_0=0$),  and since the
best fit will, by construction, follow the actual signal as closely as
possible, it necessarily overestimates the true velocity near the $v=0$
points and underestimates it near the crests.  The best-fit curve thus crosses
the true curve twice between the nodes, producing six points with zero residuals
per full period, and so an apparent 3:1 period commensuribility.

\begin{figure*}
\plottwo{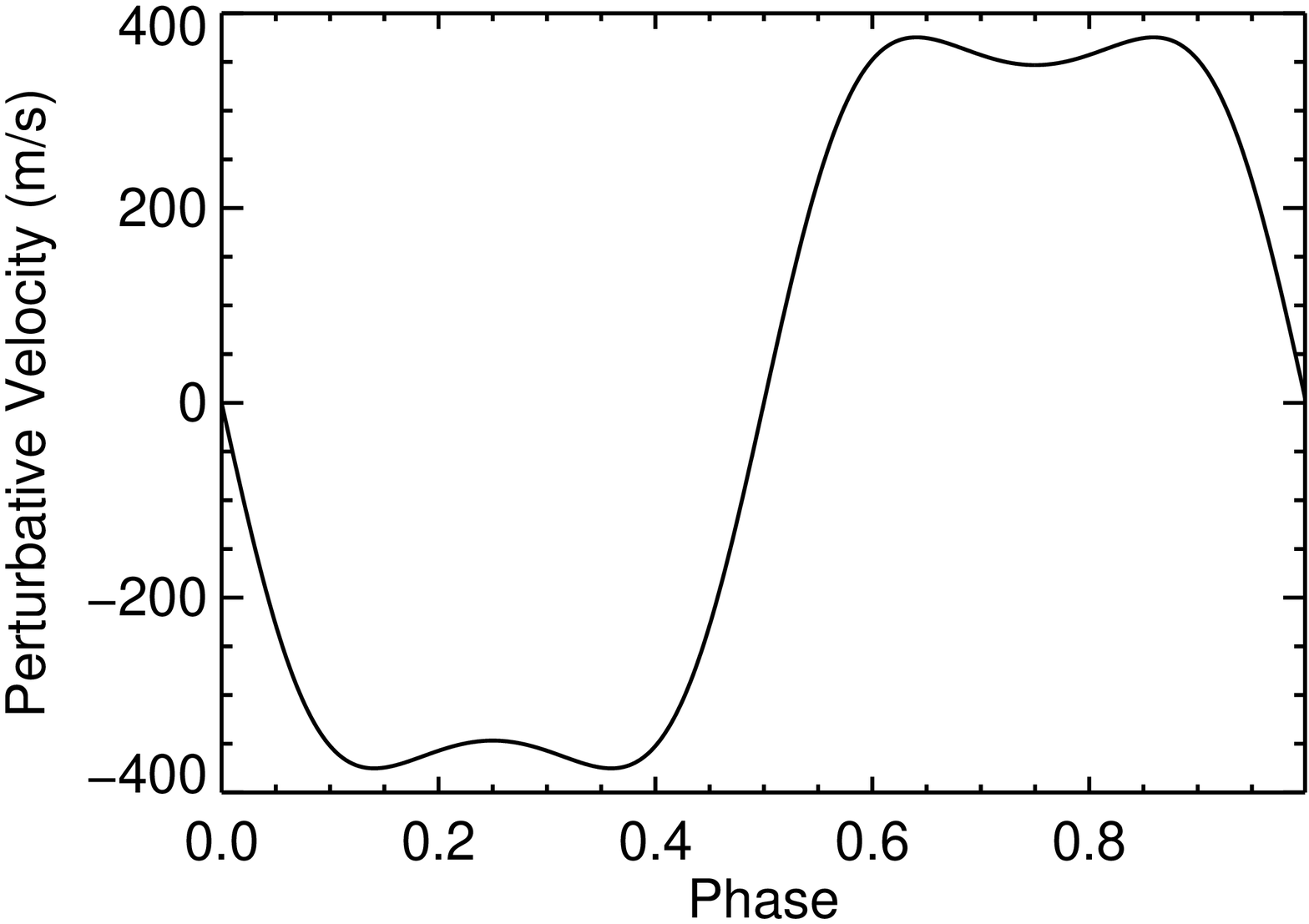}{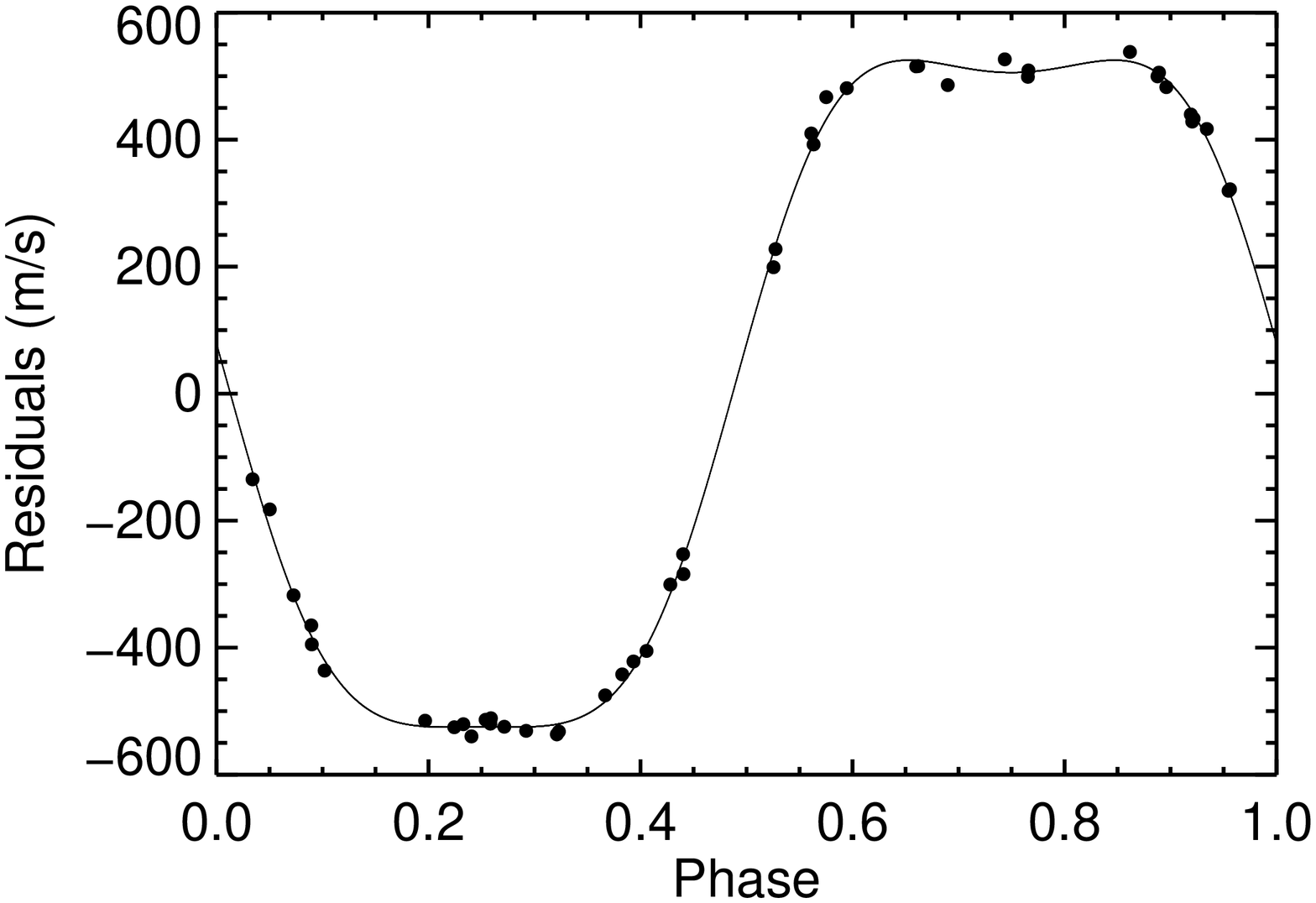}
\caption{{\it Left:} The velocity perturbations $v_{\rm{pert}}$ to the
  {\it true} RV curve  from the 
same model used in Figure~\ref{ouch}
(i.e.\ blue curve minus black curve from the left side of
Figure~\ref{ouch}, as opposed to the {\it residuals} to the {\it fit}
RV curve, which are shown in the right panel of Figure~\ref{ouch}).{\it Right:} Implied
$v_{\rm{pert}}$ curve from best fit to HET RVs of a
Keplerian-plus-``pulling''-model.  This model fits the HET data 
significantly better than the double Keplerian model.  We phase both
diagrams to have phase zero at the velocity zero with positive slope
of the primary component. \label{trueresid}}
\end{figure*}

The actual perturbative signal does not have this characteristic:  the
residuals to the {\it true} RV curve have a different characteristic
shape, as Figure~\ref{trueresid} shows. 


We have fit the HET data (as reduced with the HET template) under this scenario by assuming that the data
can be modeled as a Keplerian plus a $v_{\rm{pert}}$ component with
the free parameters given in Equation~\ref{vpert}.  The best-fit
solution is superior (r.m.s.\ $= 14$ m \persec) to that given by a double-Keplerian to the HET
data alone (r.m.s.\ $=19$ m \persec), and is consistent with the errors.  In
this solution we find $K= 3190$ m \persec, significantly higher than the fit
value for $K$ in a single-Keplerian solution, and an orbit consistent
with circular (best fit $e=0.0002 ^{+0.007}_{-0.0002}$).  The best fit perturbative
parameters are $v_0=-20$ m \persec, $k=0.30$, and $\Delta v = 4.1$ km \persec, but
are weakly constrained.  

\begin{deluxetable}{ccc}
\tablewidth{0pt}
\tablecolumns{2}\tablecaption{Best Properties of the \mbox{MARVELS-1} System\label{final}}
\tablehead{\colhead{Property} &\colhead{Value} & \colhead{Source\tablenotemark{1}}}
\startdata
Per ($d$) &         5.895322 $\pm$ 0.0002 & PPM  \\
$T_0$ (JD-2440000)  & 15509.039 $\pm$ 0.07 & PPM \\
e & 0.0002 $^{+0.007}_{-0.0002}$ & PPM \\
$\omega$ ($^\circ$) &  130 $\pm$ 0.4 & PPM \\
$K_{\rm Aa}$ (km \persec)& $3.05^{+0.01}_{-0.02}$  & CCF\\  
$K_{\rm Ab}$ (km \persec)& $3.66^{+0.06}_{-0.04}$ & CCF\\  
$q$ & $1.20^{+0.02}_{-0.01}$ & CCF\\
flux ratio & $\sim$ 0.2 & CCF\\
FWHM (Aa)  (km \persec) &  $7.26^{+0.02}_{-0.03}$  & CCF \\
FWHM (Ab)  (km \persec) & 7.31 $\pm$ 0.26 & CCF \\
i($^\circ$) & 2.47 $\pm$ 0.04 & \S~\ref{best} \\
$M_{\rm Aa}$ (\Msol) & 1.25$\pm$0.06 & \S~\ref{basics} \\
$M_{\rm Ab}$ (\Msol) & 1.04$\pm$0.05 & $M_{\rm Aa}/q$ \\
\hline
r.m.s. (m \persec) & 14.4 & PPM\\
$\chi^2_\nu$ & 0.97 & PPM \\
\enddata
\tablenotetext{1}{PPM: ``Peak pulling model'' from Section~\ref{peakpulling}.
  ``CCF'': BIS and FWHM analysis of HET spectra from Section~\ref{CCFBIS}} 
\end{deluxetable}

\subsubsection{What is the source of contamination?}

There are three obvious candidates for the origin of a contaminating
spectrum: the 0\farcs9 companion, the 0\farcs15 companion, and \mbox{MARVELS-1}
$b$.  To first order, we expect none of these to be responsible:  the
AO companions are both 4 magnitudes fainter than \mbox{MARVELS-1} in the NIR;
if they are physically associated must be lower mass than
\mbox{MARVELS-1} (and so redder) and thus contribute negligibly to the
spectrum in the optical iodine region.  \mbox{MARVELS-1} $b$, being
substellar, should have no appreciable optical emission;  indeed, any
object bound to \mbox{MARVELS-1} must have a comparable luminosity to the
star itself to be a spectral contaminant.  Nonetheless, our healthy
skepticism requires that we carefully exclude each possibility.

The Keck velocities provide a powerful diagnostic, because, unlike
with the fiber-fed HRS, the apparent velocity difference of \mbox{MARVELS-1} and its
contaminant will be a function of the position angle of slit
during the measurement if the two objects have any measurable angular
separation.  Since the Keck 0\farcs861 slit was always rotated to the parallactic
angle during our observations \citep{Filippenko82}, this would make the
signal from the 0\farcs15 companion a strong function of hour angle of
observation.  Since \mbox{MARVELS-1} was observed at hour angles ranging from
$+0.6$ to $-3.1$, this effect would make the velocity perturbations
appear essentially random, with no relationship to the phase of
\mbox{MARVELS-1} $b$.  Further, if the 0\farcs9 companion were responsible,
then its contribution would be highly variable, since its light would often miss the narrow HIRES slit entirely.

Interestingly, the Keck velocities fit the peak pulling
model equally well (r.m.s.\ $=14$ m \persec), although with a high
reduced $\chi^2$ value due to the superior internal errors.  The fact
that the slit-fed HIRES sees a nearly identical signal to the
fiber-fed HRS strongly suggests that the contaminating spectrum is
located at small angular separation, indeed within 0\farcs15. 

Although the Keck velocities show a similar signal, fitting the Keck
and HET data jointly does not produce a good fit in this model
(r.m.s.\ $=33$ m \persec), nor does fitting the HET data reduced 
with the Keck template alone (r.m.s.\ $=19$ m \persec).   This
illustrates that the form of the velocity perturbations from the
contaminating spectrum is sensitive to the choice of template
observation and the specific velocity reduction code used.  This is
perhaps not surprising, since the velocity perturbations are the
result of a failure of the forward modeling process (which assumes
uncontaminated spectra) and that different templates will have the
contaminating spectrum Doppler shifted with respect to the primary spectrum by
different amounts (at the time of the Keck template, $v=-2.0$ km
\persec; at the time of the HET template, $v=+1.3$ km \persec).

Finally, we consider the possibility that the spectrum of \mbox{MARVELS-1}
$b$ itself as a source of contamination, that is, that \mbox{MARVELS-1} is a
double-lined spectroscopic binary, and our spurious signal at 2 d is a
result of incorrect modeling of the spectrum as being only single-lined.  In this case, the RV amplitude of the
secondary spectrum could be quite large, depending on the mass ratio
of the system.  If this spectrum is contributing to Doppler
systematics, it might produce spurious (and 
small) peaks in the cross correlation function at its velocity during
the template observation, at zero velocity phase (i.e.\ during
superior and inferior conjunction, when its spectrum aligns with that
of \mbox{MARVELS-1}), and perhaps also at the velocity of any other
contaminating spectrum (e.g.\ from one of the AO companions).  

The fact that we achieve different qualities of fit using the Keck
template and the HET template is consistent with this picture, since
the velocity of the companion spectrum in the template observation
should be different between the two telescopes due to its large
intrinsic motion.  

To check this possibility more rigorously, we constructed a peak-pulling model where
the peak is pulled toward a variable velocity consistent with
that expected from \mbox{MARVELS-1} $b$, instead of a constant
velocity $v_0$.  We find that the $\chi^2$ surface to fits to the HET
data has a broad valley near its minimum, admitting significant motion
of the secondary spectrum, proportional to $\Delta v$.  Indeed, we
find that we cannot constrain the implied mass 
ratio at all, with equally good fits existing at $(\Delta v,
m_b/M) = (8.5$km \persec, 0.65$)$ and $= (3.5$ km \persec, 20.8$)$.  

The former fit corresponds to a solution where \mbox{MARVELS-1} $b$ is
stellar (that is, the system is a face-on binary, and the RV
companion is not a brown-dwarf desert object, at all).  The latter fit seems rather
unphysical, but corresponds to an essentially fixed
contaminant  (i.e.\ it is equivalent to our original
peak-pulling motion with no motion of the contaminating spectrum).  This
corresponds to a scenario in which there is a fifth object in the
\mbox{MARVELS-1}, system, a chance alignment with a star of similar
spectral type to \mbox{MARVELS-1} but only 15--30\% of its flux,
undetected in AO.

To confirm one of these or similar unlikely scenarios, we must identify intruding lines
themselves in the spectrum, and constrain whether they show
significant Doppler motion.  To break the degeneracy we find in the
peak pulling models, this analysis must involve an actual measurement
of the line widths and bisectors in the observed spectra.

\subsubsection{Contamination Estimates and a Hunt for Contaminating Lines}

We first performed a perfunctory examination of the 14 Keck spectra,
which are of high signal-to-noise ratio
and resolution, to find any obvious signs of peak pulling.  Such an
analysis would not necessarily be dispositive, because the slit-fed HIRES line
bisectors will be a strong function of the slit illumination profile,
which may overwhelm any small line profile variations intrinsic to
the source.  Nonetheless, if the variations are sufficient to induce
amplitudes of several hundred m \persec, they may be visible.

We examined two regions in the red (outside of the
iodine region):  one region with strong telluric absorption and an
adjacent order  with several strong, deep stellar lines.  We performed a
simple cross correlation in pixel space in the telluric region to
account for night-to-night lateral shifts in the spectral format, and
interpolated the spectra onto a common scale.  We then adopted an approximate
wavelength solution for the CCD (obtained from a ThAr exposure on
another night).  Next, we applied the barycentric velocity shift to
the wavelengths associated with each spectrum

We selected three moderately deep and isolated stellar lines and fit Gaussians
to their profiles, then inverted and normalized these profiles
and interpolated them into velocity (log wavelength) space with zero
velocity chosen to be the line center (as determined from the Gaussian
fit).  This approach allowed us to stack the line profiles for examination with improved signal-to-noise ratio.

\begin{figure*}
\plottwo{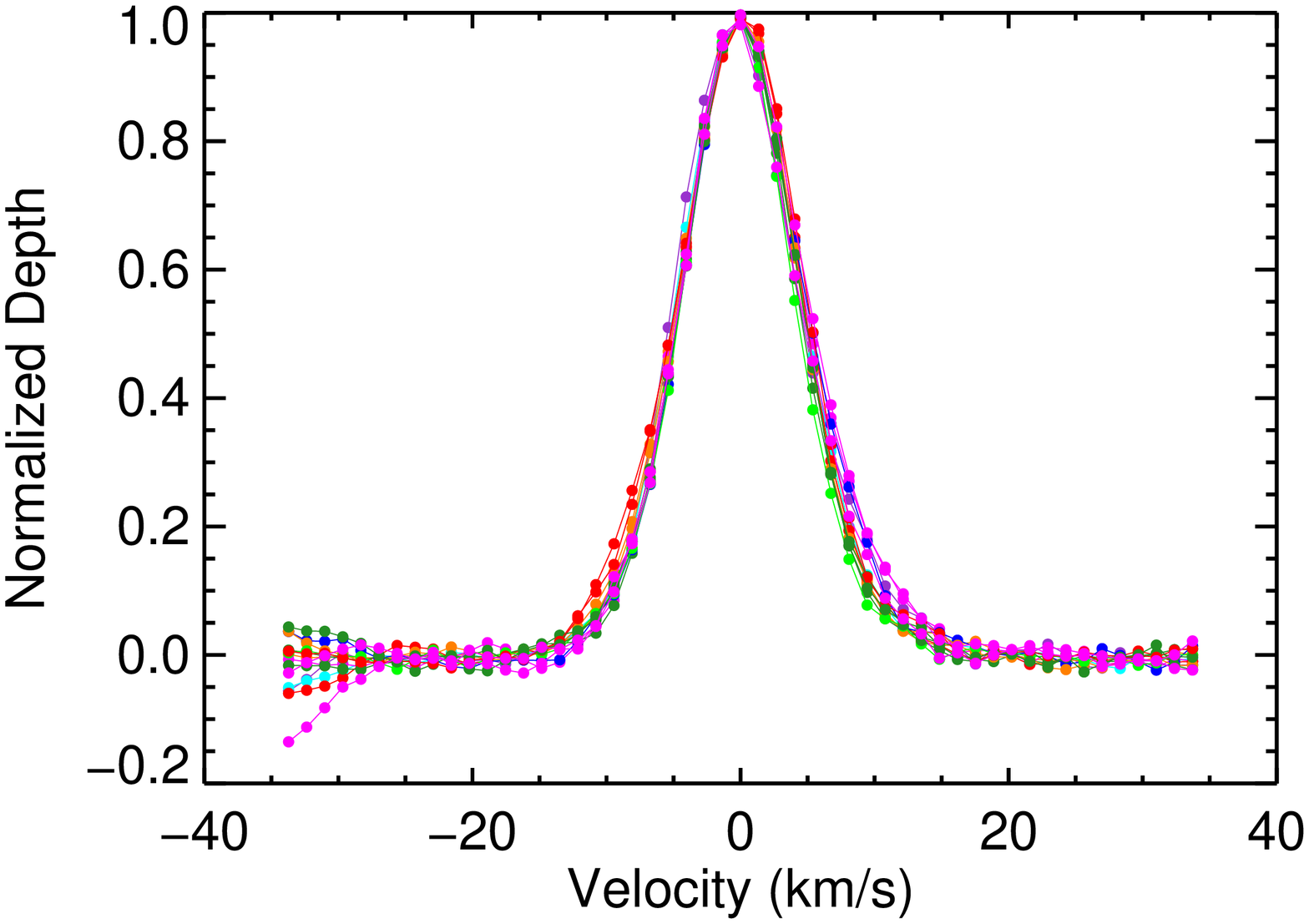}{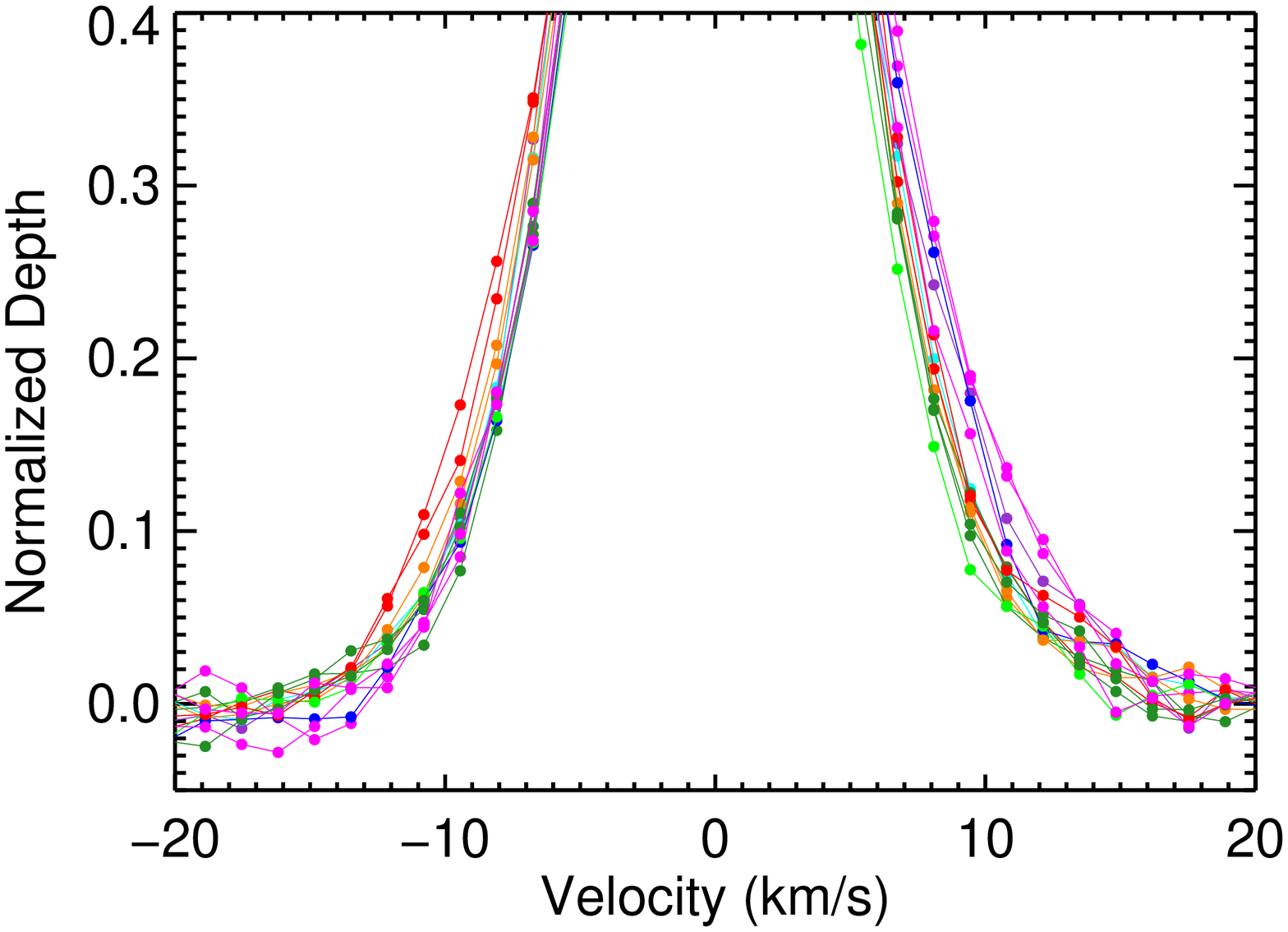}
\caption{{\it Left:} Stacked and normalized line profiles for 14
  observations of \mbox{MARVELS-1} from Keck.  Colors indicate the measured
  radial velocity of \mbox{MARVELS-1} (red indicating maximum redshift, etc.)
   {\it Right:} Detail in the line wings, showing the effect of the
   underlying contaminating spectrum.  When the star is most
   redshifted (in the barycentric frame of the \mbox{MARVELS-1} system,
   indicated by the red lines) there is a clear excess in the blue
   (negative velocity) wing of the line, and conversely on the red
   wing of the most blueshifted lines.   \label{profiles}}
\end{figure*}

We show the stacked profiles for all 14 observations in
Figure~\ref{profiles}, with each component color-coded according to the
measured radial velocity of \mbox{MARVELS-1}.  The line profile
is clearly undergoing variations, most especially in the line wings at the velocity
extremes.  This result is consistent with an underlying contaminating line at
each position with velocity at nearly exactly the average of the
line profile positions and with a slightly larger line width than that
of \mbox{MARVELS-1}.  In the frame of the figure (the frame of \mbox{MARVELS-1})
a contaminating spectrum protrudes in the wings at the
extreme velocities (emerging most clearly in the wings near $\pm 10$
km \persec).

\subsubsection{CCF Bisector analysis}
\label{CCFBIS}

Emboldened by the success of the peak pulling model at reproducing the
observed RVs and its apparently successful prediction that there would
be large line bisector variations, we proceed to perform a proper
bisector analysis of the fiber-fed HRS data, which does not suffer
from slit illumination issues. The aim of the bisector analysis was to 
measure the large line profile variations predicted by the peak pulling model, 
and thereafter recreate the correlation between orbital phase and bisector change in an attempt
to break the degeneracy between constant and variable velocity contaminants. 
In the context of planet-search surveys, there have been only a few instances where 
simulations of the bisector variation have been used to convincingly disentangle blended radial 
velocities \citep[e.g][]{Santos:2002}. Conversely, the lack of a correlation between 
RV and bisector variations has sometimes been used to constrain the parameters of a system 
or to rule out blend scenarios \citep{Torres:2004,Diaz:2012}.

Line profile bisectors have conventionally been used to distinguish between true radial velocity signatures and spectral line asymmetries that mimic the existence of a companion \citep{Toner:1988}. However, long exposure times or the stacking of many individual exposures are required in high-resolution spectroscopy to achieve the signal-to-noise ratio (SNR) required for reliable line bisector analysis, which makes it difficult to measure short-term variations. Instead, we scrutinize the bisectors of cross-correlation functions (CCFs) constructed from the cross-correlation of observed spectra with numerical line masks \citep[e.g.][]{Basturk:2011}. In addition to preserving line profile information, CCFs represent an ``average'' spectral line since they are based on multiple lines of different elements \citep{Dall:2006}. Only changes to the ensemble of spectral line profiles is reflected in the shape of the CCF, making it less vulnerable to small-scale inconsistencies.        

In an effort to study the true CCF profile, we used observations of
\mbox{MARVELS-1} on the HET HRS ``red" CCD detector (6100 - 7600 \AA) which
is beyond the iodine region. The spectra consist of 25 echelle
orders that were wavelength calibrated using the closest available
Th-Ar hollow-cathode lamp calibration exposure taken during the
night. Since the calibration exposures were neither simultaneous nor
bracketed, the wavelengths calculated by this method may suffer from
instrumental shifts. However, since we are only using these spectra
for CCF bisector analysis and not for precise radial velocity extraction, this
lack of high precision in the wavelength calibration will not effect
our analysis. Only the five most well-behaved echelle orders we used, in an 
attempt to minimize arbitrary bisector variability. We also rejected five epochs due to low 
signal to noise ratios (SNR), which produced faulty bisectors dominated by 
small-scale variation. 

We computed the CCFs for each epoch by the cross-correlation of fully
reduced and calibrated spectra with a weighted $G_2$ stellar template
mask \citep{Pepe:2002,Baranne:1996}. The mask was created by us expressly for
this purpose from an NSO FTS solar atlas
\citep{Kurucz:1984}. Consisting of 230 lines spanning the wavelength
region between 6000 - 7800 \AA, the mask has non-zero values
coincident with the wavelengths of distinct unblended stellar lines,
and reflects the relative depth of absorption lines against the local
continuum. For the cross-correlation, we use all lines that have a
depth of 5\% or more with respect to the stellar continuum, since we
lose lines by setting the mask to zero in regions where the stellar
signature is overwhelmed by telluric lines. The width of the mask
lines is adjustable, and set here to be 3 km s$^{-1}$, based
on the resolution setting of the spectrograph.  

Each spectral order was cross-correlated independently and the
resulting CCFs added to attain a composite. The bisector was 
calculated as the loci of the midpoints of horizontal lines connecting
the two wings of each composite CCF. Several measures have been used to quantify the 
shape of bisectors; we employ the bisector inverse slope (BIS), defined as
the difference of average velocities between 10-40\% and 55-85\% of
the total CCF depth \citep{Queloz:2001}. The errors on the measurement of the 
BIS and the full-width at half-maximum (FWHM) were determined  from the dispersion in each value 
between different echelle orders. All of the measured BIS and FWHM values, along with 
their $1-\sigma$ uncertainties, are presented in Table~\ref{bistab}.

\begin{figure*}[htbp] 
   \centering
      \includegraphics[width=5in]{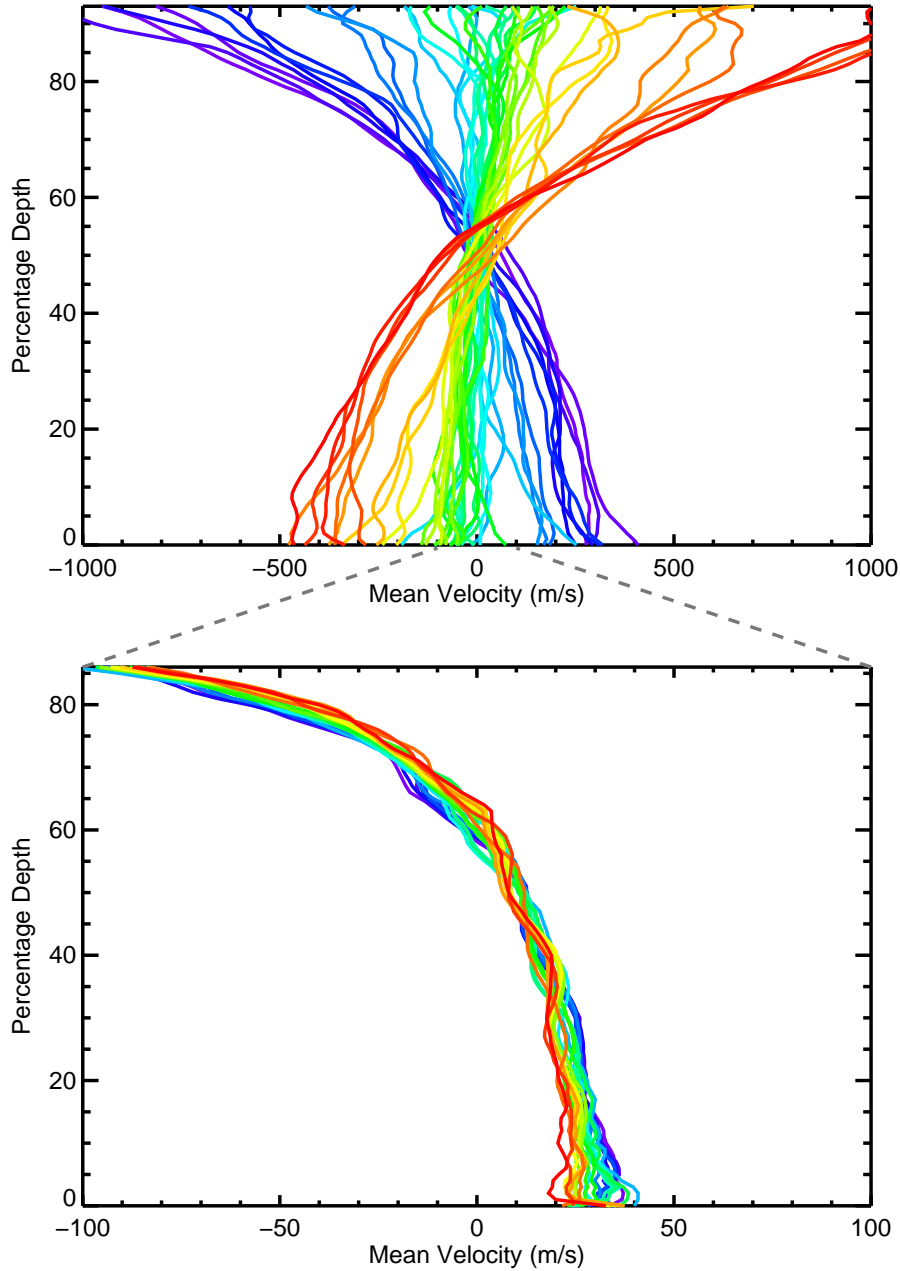} 
    \caption{\small CCF bisectors, shifted by
      their CCF-measured radial velocities to align them for clarity. The colors are based on BIS values. {\it Top:} \mbox{MARVELS-1}. These bisectors change dramatically in both shape and orientation in the period observed. {\it Bottom:} $\sigma$ Dra. Notice the factor of ten difference in the velocity axis range, demonstrating the stability of the $\sigma$ Dra bisectors.}
 \label{bis}
\end{figure*}

The bulk motion of a star would cause bisectors to oscillate around a
mean bisector without changes in shape or orientation. However, the
bisectors of \mbox{MARVELS-1} change dramatically in the period observed,
producing a large range in the BIS (Fig.\ref{bis}).  We also show observations of
$\sigma$ Dra in the same wavelength region and calibrated by the same
method for comparison; these bisectors are extremely stable and
show hardly any change. The contrast between these stars emphasizes the fact
that information about the system is encoded into the pattern of the
\mbox{MARVELS-1} bisectors.  

Figure \ref{phase}(a) shows a conspicuous relation between the BIS (measured from 
the CCF) and the cosine of the orbital phase angle, as calculated from the precise radial 
velocity results (extracted by the iodine method). Specifically, the measured velocity and 
orbital phase are related as $v=k~cos(\phi)$, where k is a given velocity semiamplitude. 
We choose to use the phase instead of 
directly using the RVs because the more precise iodine velocities are measured differently from 
the CCF method, where we simply measure the centroid of a fitted Gaussian. Even though the 
RVs produced by both methods are consistent within their respective uncertainties, we avoid 
any discrepancies by using a quantity that is agreed upon by both techniques. 

The small values of BIS near cos$(\phi)=0$ and their symmetry about that value indicates that the primary and contaminating spectra are aligned at $v=0$. Figure \ref{phase}(b) 
shows the FWHM values measured simultaneously with the BIS. 
This also has a minimum near cos$(\phi)=0$, reaffirming that the spectra are aligned at $v=0$.
This is consistent with the result from our peak pulling model that $v_0\sim0$.

\subsubsection{A Two-Component CCF model}

We attempted to reproduce this relation between BIS and cos$(\phi)$ with a
simple model. In keeping with the idea that spectral
contamination from an unseen companion is leading to ``peak pulling'',
we simulated a scenario where a large Gaussian ($G_1$, representing the
true CCF of \mbox{MARVELS-1} alone) shifts relative to a secondary Gaussian
($G_2$, representing the hypothesized contaminant), and periodically aligns with
it at some velocity shift. At each phase step we calculated the BIS and FWHM of the combined
CCF, produced by the sum of $G_1$ and $G_2$. A similar simulation with two Gaussians was 
performed by \cite{Santos:2002} to recreate a linear BIS-RV relationship. They were able to 
detect a potential brown dwarf around a member of a binary system, using a constant velocity 
$G_1$ and four free parameters (including a variable $G_2$). 

There are six free parameters in our model: the ratio 
of amplitudes of the two Gaussians ($A_2/A_1$), the FWHM of $G_1$ 
($w_1$), the FWHM of $G_2$ ($w_2$), the mass ratio ($q=M_2/M_1$), 
the velocity semiamplitude of $G_1$ ($K_1$), and an adjustable offset between
measured model BIS and observed BIS ($\Delta_{\rm BIS}$). This offset is included to allow
for the fact that the observed CCF is not a perfect Gaussian, as assumed in the model.
We explored the possibility of a constant velocity contaminant with this model, and found 
that it was not possible to successfully match both BIS and FWHM with this configuration, 
although the BIS could be reproduced by itself. Thus, the behavior of the line profiles 
rules out the presence of a fifth object in chance alignment with this system. 

\begin{figure}[htbp] 
   \centering
      \includegraphics[width=3.2in]{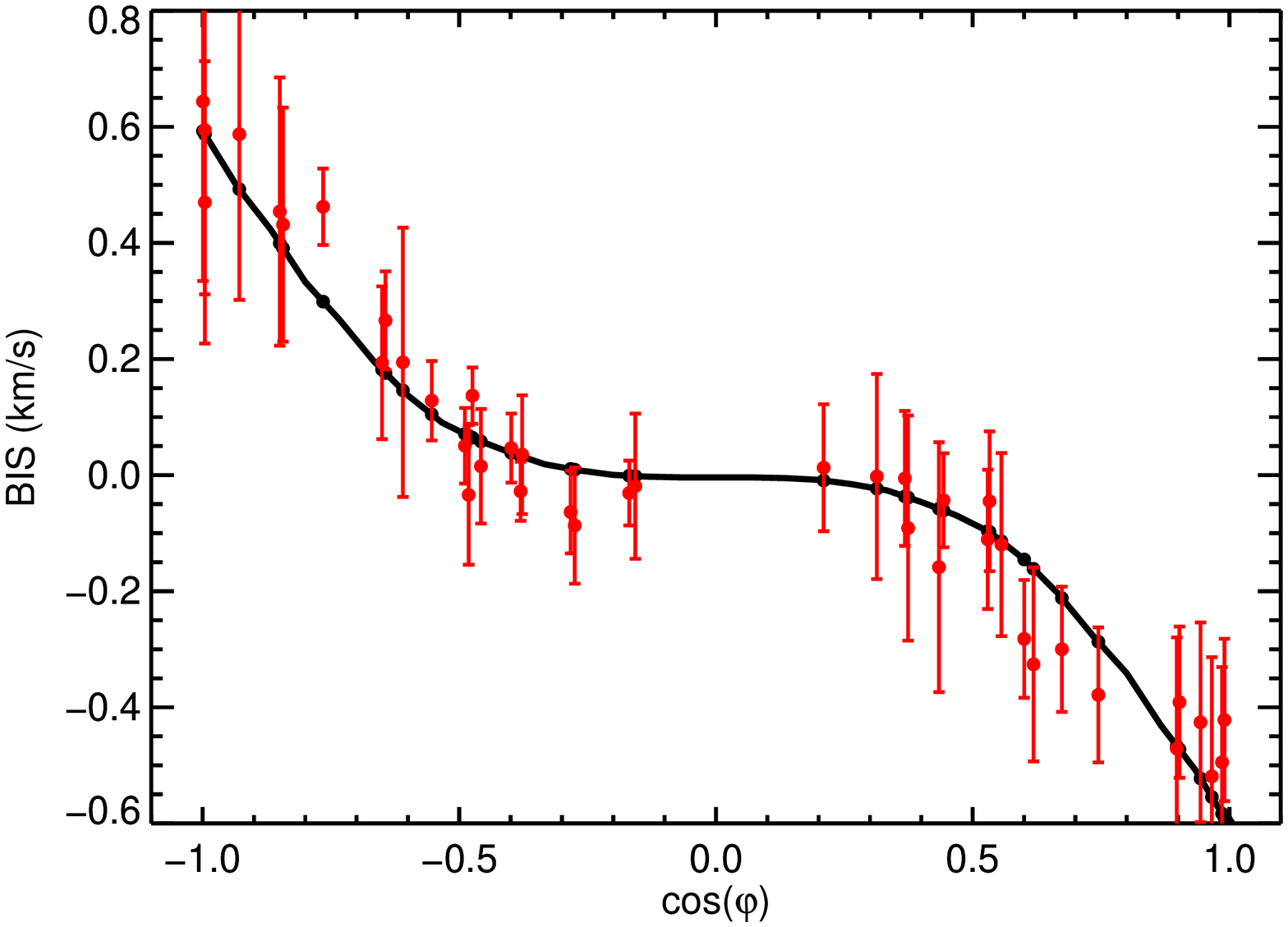} 
      \includegraphics[width=3.2in]{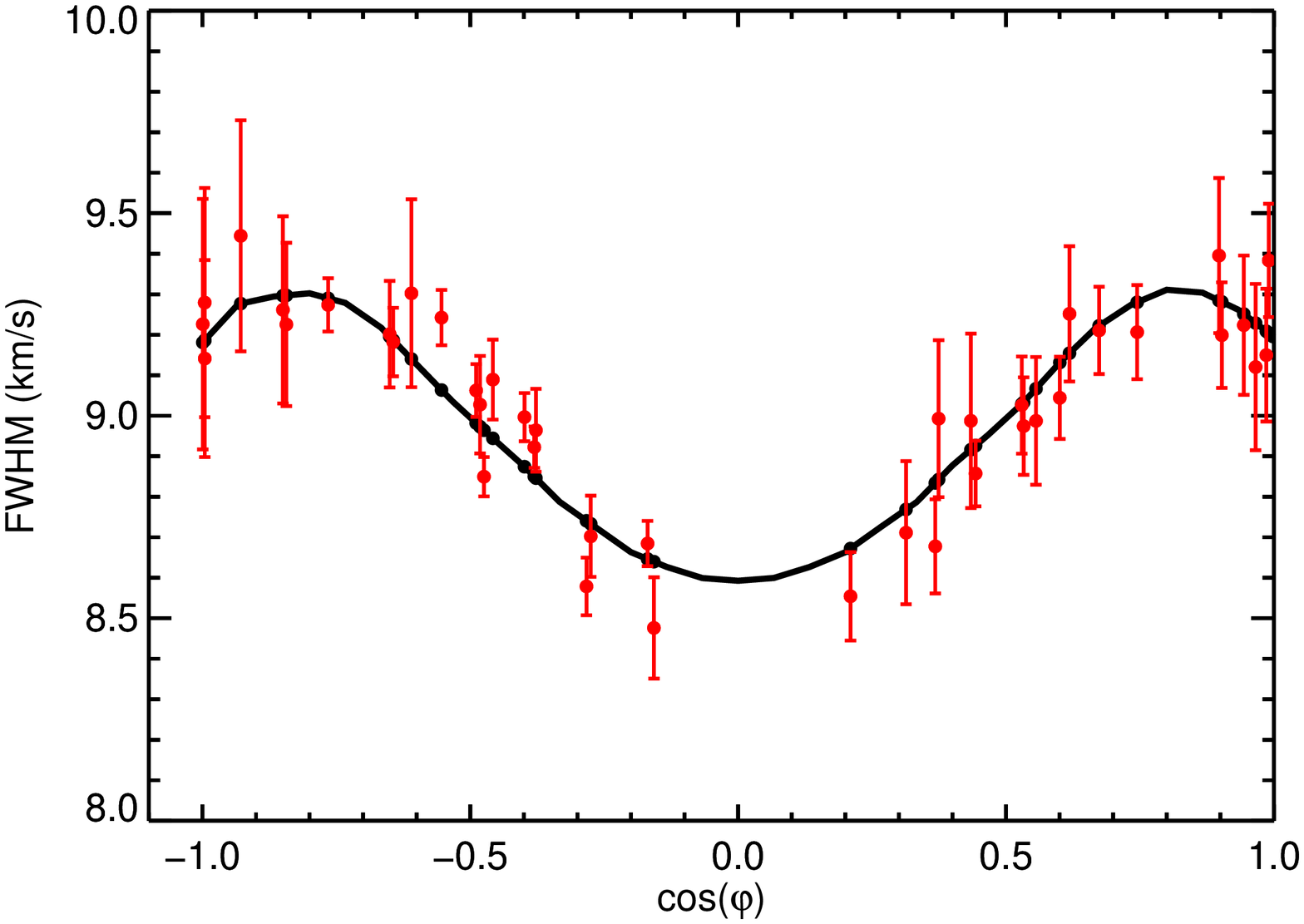} 
\caption{\small {\it(a, left)} Points showing the relation between
    observed CCF bisector (BIS) values and the cosine of the measured phase 
     angle (cos$\phi$) for \mbox{MARVELS-1} (for convenience, the phase here is defined
     such that phase zero ($\varphi=0$) occurs at the maximum radial
     velocity for the best-fit circular orbit). The solid line shows the BIS 
     calculated from the simple model consisting of two shifting Gaussians, using parameters chosen to
     fit the observed BIS and full-width at half-maximum (FWHM) values
     as a function of this orbital phase. {\it(b,right)} Points as in {\it (a)}; line shows the full-width at
     half-maximum (FWHM) calculated simultaneously from the two-Gaussian model. In both plots the
     bisector inverse spans and widths of the model and observed CCFs were measured
     in an identical fashion. The error bars are determined from the measurement variation between echelle orders.}
  \label{phase}
\end{figure}

A variable velocity contaminant renders an extremely viable model.
In this scenario, the Gaussians undergo motion as in a binary system with a 
given mass ratio. In other words, the velocity semiamplitude of $G_2$ is determined by the mass 
ratio and the motion of $G_1$, or $K_2 = - q K_1$. Figure
\ref{phase}(a) shows the result of our simulation, where both BIS and
FWHM are matched {\it simultaneously} by mimicking a binary system. We
arrive at our best fit parameters using an AMOEBA routine \citep{AMOEBA}, and
conclude that the observations are modeled well by a binary system
with $q = 1.20^{+0.02}_{-0.01}$, $K_1 = 3.05^{+0.01}_{-0.02}$ km
s$^{-1}$, $A_2/A_1 = 18.93^{+0.83}_{-0.64}\%$, $w_1 =
7.26^{+0.02}_{-0.03}$ km s$^{-1}$, $w_2 = 7.31^{+0.26}_{-0.26}$ km
s$^{-1}$, $\Delta_{\rm BIS} = -0.11\pm 0.02$ km s$^{-1}$.  Errors
quoted here are formal errors on the parameters, which do not include
systematic effects or covariances with the other parameters.  Note that $A_2/A_1$
here is a proxy for the flux ratio of the binary.  These formal uncertainties
here were calculated by fixing all parameters, but one which was
allowed to float, and determining the value of the floating parameter
that increased $\chi^2$ by 1 above its minimum value (30.5, with 34 degrees
of freedom).

The fact that we are able to model the line profile variability so convincingly suggests
that the culprit for the observed anomaly in the \mbox{MARVELS-1} RV
residuals is in fact a massive, relatively bright, bound companion.

\begin{deluxetable*}{ccccc}
\tablecolumns{5} 
\tablewidth{0pt}
\tablecaption{Observed values of BIS and FWHM for HET HRS data of \mbox{MARVELS-1} \label{bistab}} 
\tablehead{\colhead{time} & \colhead{BIS} & \colhead{Uncertainty$_{\rm BIS}$} & \colhead{FWHM} & \colhead{Uncertainty$_{\rm FWHM}$}\\
 \colhead{BJD$-$2440000 (UTC)} &
  \colhead{m \persec} & \colhead{m \persec} & \colhead{m \persec} & \colhead{m \persec}}

\startdata 
          15177.6109 &       0.06 &   0.1 &       8.97 &   0.1\\
          15178.5997 &       0.14 &   0.1 &       8.96 &   0.1\\
          15180.7994 &       0.15 &   0.1 &       9.00 &   0.1\\
          15181.7885 &      -0.01 &   0.1 &       9.03 &   0.1\\
          15182.7878 &      -0.37 &   0.2 &       9.40 &   0.2\\
          15183.5788 &      -0.06 &   0.2 &       8.99 &   0.2\\
          15184.5782 &       0.12 &   0.1 &       9.09 &   0.1\\
          15185.5711 &       0.75 &   0.3 &       9.23 &   0.3\\
          15483.7415 &      -0.20 &   0.1 &       9.21 &   0.1\\
          15483.7528 &      -0.22 &   0.2 &       9.25 &   0.2\\
          15484.9464 &       0.08 &   0.1 &       8.48 &   0.1\\
          15484.9577 &       0.07 &   0.1 &       8.68 &   0.1\\
          15485.7397 &       0.53 &   0.2 &       9.23 &   0.2\\
          15485.7510 &       0.56 &   0.2 &       9.26 &   0.2\\
          15497.7059 &       0.69 &   0.3 &       9.44 &   0.3\\
          15498.7192 &       0.57 &   0.1 &       9.27 &   0.1\\
          15498.9215 &       0.30 &   0.2 &       9.30 &   0.2\\
          15500.6947 &      -0.28 &   0.1 &       9.21 &   0.1\\
          15500.9073 &      -0.32 &   0.2 &       9.22 &   0.2\\
          15501.6961 &      -0.18 &   0.1 &       9.04 &   0.1\\
          15507.6863 &      -0.02 &   0.2 &       8.99 &   0.2\\
          15510.6632 &       0.30 &   0.1 &       9.20 &   0.1\\
          15510.6715 &       0.37 &   0.1 &       9.18 &   0.1\\
          15522.6391 &       0.15 &   0.1 &       9.06 &   0.1\\
          15522.6469 &       0.07 &   0.1 &       9.03 &   0.1\\
          15522.6547 &       0.24 &   0.0 &       8.85 &   0.0\\
          15522.8492 &       0.04 &   0.1 &       8.58 &   0.1\\
          15522.8570 &       0.02 &   0.1 &       8.70 &   0.1\\
          15523.6459 &       0.06 &   0.1 &       8.86 &   0.1\\
          15524.6390 &      -0.39 &   0.2 &       9.15 &   0.2\\
          15527.6284 &       0.57 &   0.2 &       9.14 &   0.2\\
          15576.7008 &       0.01 &   0.2 &       8.99 &   0.2\\
          15577.7000 &      -0.32 &   0.1 &       9.38 &   0.1\\
          15777.9371 &      -0.29 &   0.1 &       9.20 &   0.1\\
          15779.9355 &       0.08 &   0.1 &       8.92 &   0.1\\
          15782.9385 &       0.10 &   0.1 &       8.68 &   0.1\\
          15783.9280 &      -0.42 &   0.2 &       9.12 &   0.2\\
          15790.9292 &       0.10 &   0.2 &       8.71 &   0.2\\
          15791.9106 &       0.23 &   0.1 &       9.24 &   0.1\\
          15792.9209 &       0.70 &   0.3 &       9.28 &   0.3\\
          15796.8955 &       0.12 &   0.1 &       8.55 &   0.1\\
\enddata
\end{deluxetable*}

\subsubsection{Final best parameters for the MARVELS-1 binary system}

\label{best}
The mass ratio and amplitude of the
RV primary component should be well represented by the $q$ and $K_1$
parameters in our two-component CCF model.
Interpretation of $A_2/A_1$ as the flux ratio of the two stars is
complicated by that quantity's dependence on the choice of lines used
in the two models; since we have not extracted the spectrum of 
the secondary here we should anticipate that this is only a rough
estimate of the true flux ratio.    

The widths of the Gaussians ($w_1, 
w_2$) should correspond to the actual FWHM of the lines in the
individual spectra (these values were computed after subtraction of the
instrumental profile), and would indicate an equatorial rotational velocity $v \sin{i} =
\mbox{FWHM}/\sqrt{3} \sim 4$ km \persec\ if we assume that a rotation
broadening kernel of a uniform intensity disk with no differential
rotation dominates the line width.  We have validated this estimate by applying
our CCF code to synthetic spectra and the Solar spectrum convolved with
rotational kernels or various equatorial velocities to calibrate our
method.  We find that for the FWHM values we measure are consistent
with $v\sin{i}$ values of 3.5--4.5 km \persec, favoring the higher values.

The parameter $\Delta_{\rm BIS}$ encapsulates
information about the inherent bisector span of the blended spectrum
when the two spectra are perfectly aligned in velocity space.  In any
case, the formal uncertainty on these model parameters underestimates the
true uncertainties in the physical parameters of the system because of
the crudeness of our model of stellar spectral CCFs as pure Gaussians.

We calculate the inclination of the binary orbit from the binary mass
function we infer from the CCF model:
\begin{equation}
\sin^3{i} = \frac{K_1^3 q (1+q)^2 P}{2 \pi G M_{\rm Aa}}
\end{equation}
\noindent where $M_{\rm Aa}$ is the mass of the primary star.
Propagation of errors using the formal uncertainty on $q$ and $K$ and
assuming $M_{\rm Aa} = 1.25 \pm 0.06 \Msol$ (Section~\ref{basics})
yields $i = 2\fdg47 \pm 0\fdg04$, where the uncertainty in $i$ is
dominated by the uncertainty in the mass of the primary.

We report our best system parameters in Table~\ref{final}.

\subsection{Summary of Potential False Alarm Consideration}
\label{FAsummary}

We have found that the combination of the 3:1 period commensuribility of the
signals, their relative and absolute strengths, and the stability of
the photometry, rule out all scenarios that we have considered but
three: a planetary companion in a 3:1 mean motion resonance, a binary brown
dwarf in a 3:2 orbit-orbit resonance, and a 15-- 30\% spectral contaminant with
a systemic velocity nearly identical to that of \mbox{MARVELS-1} from an
unresolved companion.  

Our line profile analysis definitively identifies \mbox{MARVELS-1}$b$
as the source of the contaminating spectrum, allowing
us to dispose of the other hypotheses.  The high inclination of this
nearly-face-on binary dilutes the true Doppler signature of the orbit
down to values consistent with a brown-dwarf-desert candidate, and the
effects of the contaminating spectrum create residuals to a sinusoidal
fit that mimic a 3:1 resonance.

\subsection{Similarities with Transit Searches}
\label{rarity}

How can we understand why such a rare face-on binary system happens to be the first
system detected by MARVELS?  Fortunately, we can achieve this understanding without a
thorough analysis of the MARVELS selection effects, which is far
beyond the scope of this work.

MARVELS has observed a total of 3,300 FGK
stars with V=7.6-12 (a magnitude limited selection) in 2008-2012 with each observed about 27 times
over 2 year window.   In contrast, other radial velocity planet
searches have typically searched of order 1,000 stars.  Because of its
design, MARVELS necessarily targets 
stars whose multiplicity, variability, and spectra properties are less
well-known beforehand.  In addition, because the target stars are
fainter, they are mostly likely to be blended with a significant
contaminant (both because they are fainter and because they are more
distant, and so there is a higher probability of a blended companion).
MARVELS is thus especially sensitive to blend
scenarios and binaries, relative to other RV planet searches, and needs to confirm
its lowest amplitude detections in a manner similar to transit searches.

In a similar vein, the MARVELS program has also recently discovered
the double-lined highly eccentric spectroscopic binary 
TYC-3010-1494-1, which masquerades as a modest-eccentricity brown
dwarf candidate (Mack et al. 2013, AJ, accepted).  
 
Nonetheless, the probability of the particular blend scenario of
MARVELS-1 seems to low for such an object to be expected in the
MARVELS survey.   A rough order of magnitude calculation would
estimate a $\sim 50\%$ binary fraction for stars, that $\sim 10\%$ of
binaries are close binaries, that perhaps $\sim 10\%$ have mass ratios
such that the secondary would contaminate the spectrum, and that
$0.1\%$ of systems would be within 3 degrees of face-on (so the RV
amplitude of the orbit is comparable to the line widths).  This would 
suggest that MARVELS should encounter
$3300\times0.5\times0.1\times0.1\times0.001\sim 0.02$ such systems.
Either these estimates are collectively off by two orders of magnitude, or the
MARVELS survey was quite unlucky to have encountered this system.

This situation is in some ways similar to that of transit searches,
which have required extensive efforts to confirm and validate planet
candidates, which must be sifted out from a large number of classes of
false positive, in particular blends with eclipsing binary stars.  For
instance, \citet{Mandushev05} found a similarly insidious form of 
false positive in the form of a eclipsing binary orbiting a slightly
evolved F star that produced a nearly achromatic photometric signal indistinguishable
from a transit in their photometric wide-field transit survey, and
produced a similar ``peak pulling'' signal to the one we see here. 

\section{Nomenclature}

\label{nomenclature}
Nomenclature standards for multiple systems are subjective and tricky
to apply consistently.  If we
attempt to follow the Washington Multiplicity Catalog (WMC) standard as
described in \citet{Raghavan2010} and recommended by the International
Astronomical Union (IAU), then we must still decide how to
organize these (possibly unbound) companions hierarchically and in
what order to assign component letters.   

The natural hierarchy here is unclear, because we might designate the 0\farcs9 companion ``B'', and to group the other three
components into a nearly-unresolved ``A''  system, or we might put the
0\farcs15 and 0\farcs9 companions on equal footing and dub them ``B''
and ``C'' respectively.  We choose the latter (consistent with our
Figure~\ref{Ji}).  This makes the face-on short period binary \mbox{MARVELS-1} Aa and
\mbox{MARVELS-1} Ab (where the lowercase letters follow the WMC
convention for stars, but are conveniently consistent with the prior
names for these objects, \mbox{MARVELS-1} and \mbox{MARVELS-1} $b$,
respectively).    We note that \mbox{MARVELS-1} C is unlikely to be bound.



\section{Conclusions}
\label{conclusions}

We have performed a thorough analysis of the complex \mbox{MARVELS-1} system.  Using
adaptive optics imaging, we have identified a companion at 0\farcs9 separation that
appears to be a foreground late M dwarf, and another at 0\farcs15 separation that
appears to be a late K or early M dwarf associated with the primary, which we classify
as a F9 dwarf.

The primary was previously thought to host a short-period companion
occupying the ``brown-dwarf desert'', \mbox{MARVELS-1} $b$, the first sub-stellar companion
discovered with the MARVELS instrument.  Follow-up radial velocity measurements
revealed strong deviations from a Keplerian solution, with amplitude
$\sim 100$ m \persec and a an orbital frequency exactly three times
that of \mbox{MARVELS-1} $b$.

We have identified three extraordinary explanations for the observed
radial velocity signature of \mbox{MARVELS-1}:  a pair of substellar objects
in a near-perfect 3:1 mean-motion resonance with strong dynamical interactions, a binary
brown dwarf in a 3:2 resonance, and a blended stellar component
contaminating the spectra, creating apparent residuals
to a Keplerian solution at three times the observed period.  

Identification of strong line bisector variations consistent with a
contaminating spectrum confirms that the final scenario as the correct
one, and our detection of significant radial velocity motion of the
contaminating spectrum confirms that it is due, in fact, to
\mbox{MARVELS-1} $b$ itself, which is actually a stellar binary
companion in a face-on orbit.

In this case, the concern raised over the unusual properties of the
system caught an unlikely and insidious form of spectral
contamination.  A routine check for line profile variations would have
caught the problem early, but there is little motivation for such a
system at Keck, where large line profiles variations are expected,
even on RV stable stars, due to changes in the spectrograph and slit
illumination.   Bisector analysis is further hampered on unstabilized
spectrographs by the lack of a precise wavelength scale outside the
iodine region, and by the presence of iodine lines within the iodine region.
Only in cases such as MARVELS-1 with large (3 km \persec) RV
variations, or with fiber-fed spectrographs, can bisector changes be obviously attributed to
astrophysical, as opposed to instrumental, effects.  Nonetheless, such
cases may be common among hot Jupiters discovered with  {\it Kepler},
and in such cases a line bisector analysis from Keck spectra will be
an important validation step.

In the absence of such a check, the only indication that something was amiss was a ``feeling'' that the system 
was too unusual (the resonance was too perfect, the two-planet fit
never good enough, the phases of the two ``planets'' too well matched)
and the difficulty in identifying a good dynamical solution.  

In summary, \mbox{MARVELS-1} appears to be a stellar triple, with one presumably
bound companion seen in AO and the other detected by radial velocities
and line bisector variations.  A foreground apparent companion
is separated from this system by 0\farcs9 on the sky, bringing the total number of
detected companions to the primary object to three.

\acknowledgments

We thank the referee, Dr.\ Alexandre Santerne, for a thorough
and thoughtful review, and most especially his encouragement to
more rigorously explore the possibility that \mbox{MARVELS-1} is face-on binary.

Funding for SDSS-III has been provided by the Alfred P. Sloan Foundation, the Participating Institutions, the National Science Foundation, and the U.S. Department of Energy. The SDSS-III web site is http://www.sdss3.org/.

SDSS-III is managed by the Astrophysical Research Consortium for the Participating Institutions of the SDSS-III Collaboration including the University of Arizona, the Brazilian Participation Group, Brookhaven National Laboratory, University of Cambridge, University of Florida, the French Participation Group, the German Participation Group, the Instituto de Astrofisica de Canarias, the Michigan State/Notre Dame/JINA Participation Group, Johns Hopkins University, Lawrence Berkeley National Laboratory, Max Planck Institute for Astrophysics, New Mexico State University, New York University, Ohio State University, Pennsylvania State University, University of Portsmouth, Princeton University, the Spanish Participation Group, University of Tokyo, University of Utah, Vanderbilt University, University of Virginia, University of Washington, and Yale University.

NSO/Kitt Peak FTS data used here were produced by NSF/NOAO.

The Center for Exoplanets and Habitable Worlds is supported by the
Pennsylvania State University, the Eberly College of Science, and the
Pennsylvania Space Grant Consortium.

This work was supported by the NASA Astrobiology Institute through the
Penn State Astrobiology Research Center (grant NNA09DA76A).

We acknowledge the University of Florida High-Performance
Computing Center for providing computational resources and support
that have contributed to the results reported within this paper.  This
research has made use of NASA's Astrophysics Data System.  This
research has made use of the SIMBAD database, operated at CDS,
Strasbourg, France 

The Hobby-Eberly Telescope (HET) is a joint project of the University of Texas at Austin, the Pennsylvania State University, Stanford University, Ludwig-Maximillians-Universit\"at M\"unchen, and Georg-August-Universit\"at G\"ottingen. The HET is named in honor of its principal benefactors, William P. Hobby and Robert E. Eberly.

We thank the Penn State HET Science Director for the
allocation of discretionary time necessary for this work, and the Penn
State HET TAC for allocating regular time for the followup of MARVELS
targets, including those described herein.

We thank NASA and NExScI for providing Keck time in the 2011B semester
for the study of multiplanet systems (NExScI ID40/Keck ID\# N141Hr, PIs
Wright \& Ford).  This work was supported by a NASA Keck PI Data
Award, administered by the NASA Exoplanet Science Institute. Data
presented herein were obtained at the W. M. Keck Observatory  from
telescope time allocated to the National Aeronautics and Space
Administration through the agency's scientific partnership with the
California Institute of Technology and the University of
California. The Observatory was made possible by the generous
financial support of the W. M. Keck Foundation.  We thank the
California Planet Survey consortium, 
and especially Andrew Howard and Geoff Marcy, for managing the queue,
undertaking the Keck radial velocity measurements and template
observation, and the precise Doppler analysis.  

We thank Debra Fischer for use of her precise Doppler pipeline for the
early stages of this effort, for encouraging a more thorough analysis
of potential false positives, and for key insights into the pernicious
nature of spectral contamination. 

J.T.W.\ performed much of the analysis not mentioned below, oversaw
and managed the overall research effort, and prepared this
manuscript, including the incorporation of figures and text provided
by the other authors.  A.R., under the direction of J.T.W.\
and S.M., performed the CCF and line bisector and CCF analysis, and
provided the crucial proof that \mbox{MARVELS-1} is a double-lined
binary.  J.T.W.\ and S.M.\ proposed for, obtained, and  
executed the HET observations, and developed most of the false
positive analysis.  
S.X.W.\ performed the raw reduction of the HET spectra and the precise
Doppler analysis.  M.J.P. performed the MCMC and DEMCMC dynamical 
analysis and stability tests and TTV calculations with guidance from
E.B.F. who led the development of the MCMC, DEMCMC and TTV algorithms
used for the analysis.  J.C.\ obtained
the AO imaging and he and Ji W.\ 
performed the image analysis and derived astrometry and photometry of
the \mbox{MARVELS-1} companions.   B.S.G.\ and
J.P.\ supplied the KELT photometry and provided its analysis.  J.G.\
is PI of MARVELS and provided encouragement and insight into the
nature of the \mbox{MARVELS-1} system and the MARVELS instrument.
John W., B.L.L., P.A.C., J.I.G.H., L.G., L.D-F., G.F.P.M., M.A.G.M, L.N.C,
R.L.C.O, and B.X.S.\  developed the stellar characterization pipeline
and derived the new stellar parameters from high resolution spectra
based on a reanalysis of the FEROS spectra and new analysis of the APO
spectra.  John W.\ proposed for, observed, and reduced the APO
spectra.  Other 
authors provided significant contributions to this research through
their membership in the SDSS consortium and/or the MARVELS team, and
with useful comments on this paper.  

E.B.F.\ and M.J.P.\ were supported by NASA Origins of Solar
Systems grant NNX09AB35G.

Work done by B.S.G.\ and J.E.\ was supported by NSF CAREER grant AST-1056524.

J.P.\ and K.G.S.\ were supported by the Vanderbilt Office of the Provost through
the Vanderbilt Initiative in Data-intensive Astrophysics (VIDA).  The
SME work at Vanderbilt was sponsored through the NSF Astronomy \&
Astrophysics grant AST-1109612. 
 
Funding for the Brazilian Participation Group has been provided by
the Minist\'erio de Ci\^encia e Tecnologia (MCT), Funda\c c\~ao Carlos
Chagas Filho de Amparo \`a Pesquisa do Estado do Rio de Janeiro
(FAPERJ), Conselho Nacional de Desenvolvimento Cient\'{\i}fico e
Tecnol\'ogico (CNPq), and Financiadora de Estudos e Projetos (FINEP).

G.F.P.M. acknowledges financial support by CNPq (476909/2006-6 and
474972/2009-7) and FAPERJ (APQ1/26/170.687/2004) grants.

L.G.\ acknowledges financial support provided by the PAPDRJ
CAPES/FAPERJ Fellowship.  L.D.F. acknowledges a PhD scholarship from CAPES/Brazil.

\facility{{\it Facility} Keck:I}


\end{document}

%% file: tab1.tex
15175.5879 & 623.88 & 17.6\\
15177.6109 & 1391.46 & 17.5\\
15178.5997 & -979.18 & 16.6\\
15180.7994 & -1038.28 & 17.1\\
15181.7885 & 1291.27 & 16.8\\
15182.7878 & 2433.65 & 17.4\\
15183.5788 & 1186.79 & 15.1\\
15184.5782 & -1170.84 & 18.1\\
15185.5711 & -2840.64 & 14.6\\
15483.7415 & 2159.31 & 13.5\\
15483.7528 & 2142.75 & 14.7\\
15484.9464 & -489.47 & 13.5\\
15484.9577 & -507.08 & 12.7\\
15485.7397 & -2353.73 & 14.4\\
15485.7510 & -2359.49 & 17.4\\
15497.7059 & -2631.09 & 18.7\\
15498.7192 & -2078.22 & 15.5\\
15498.9215 & -1623.45 & 20.6\\
15499.7351 & 359.36 & 16.4\\
15500.6947 & 2311.46 & 18.1\\
15500.9073 & 2468.96 & 17.6\\
15501.6961 & 1699.77 & 18.4\\
15507.6863 & 1526.22 & 14.1\\
15510.6632 & -1762.64 & 15.0\\
15510.6715 & -1711.09 & 15.5\\
15522.6391 & -1285.58 & 18.2\\
15522.6469 & -1283.72 & 16.5\\
15522.6547 & -1258.66 & 17.9\\
15522.8492 & -777.53 & 17.7\\
15522.8570 & -728.73 & 15.8\\
15523.6459 & 1120.44 & 13.0\\
15524.6390 & 2472.26 & 17.2\\
15524.8380 & 2341.07 & 18.8\\
15527.6284 & -2843.01 & 15.2\\
15531.6080 & 708.58 & 17.2\\
15576.7008 & 1109.81 & 21.7\\
15577.7000 & 2483.02 & 16.9\\

%% file: tab2.tex
15769.0930 &      -2349.90 &      5.6 \\
  15782.1024 &      -680.13 &      4.3 \\
  15783.1275 &       1666.41 &      5.2 \\
  15789.1125 &       1886.25 &      5.5 \\
  15791.1307 &       838.79 &      3.3 \\
  15793.1094 &      -2244.26 &      5.4 \\
  15796.1305 &       2644.81 &      5.9 \\
  15797.1316 &       599.29 &      2.8 \\
  15798.0922 &      -1702.91 &      5.5 \\
  15808.0954 &       2335.50 &      6.5 \\
  15809.0594 &       266.61 &      3.1 \\
  15810.0965 &      -2077.35 &      5.6 \\
  15811.0929 &      -1821.60 &      4.6 \\
  15812.1337 &       643.32 &      3.6 \\